\documentclass[review]{elsarticle}

\usepackage{setspace}
\usepackage{amssymb}
\usepackage{amsmath,graphicx,hyperref}
\usepackage{booktabs}
\usepackage{url}
\usepackage{tcolorbox}
\usepackage{tabularx}
\usepackage{multirow}

\usepackage{lineno}

\journal{Signal Processing}

\begin{document}

\begin{frontmatter}


\title{Spatial-Frequency Cued Generative Fixed-Filter Active Noise Control Based on Deep Learning in Reverberant Environments}


\author[ntu]{Boxiang Wang} 
\author[ntu]{Haowen Li} 
\author[npu]{Dongyuan Shi} 
\author[ntu]{Junwei Ji} 
\author[ntu]{Ziyi Yang} 
\author[ntu]{Zhengding Luo}
\author[ntu]{Woon-Seng Gan} 

\affiliation[ntu]{organization={Smart Nation TRANS Lab},
            addressline={School of Electrical and Electronic Engineering, Nanyang Technological University}, 
            city={Singapore},
            postcode={639798}, 
            country={Singapore}}
 
\affiliation[npu]{organization={Center of Intelligent Acoustics and Immersive Communications},
            addressline={Northwestern Polytechnical University}, 
            city={Xi'an},
            postcode={710071}, 
            country={China}}


\begin{abstract}
Generative fixed-filter active noise control (GFANC) effectively attenuates noise with diverse frequency characteristics through the combination of sub control filters. However, it does not incorporate the spatial information of the noise source, which limits its performance, particularly in reverberant environments. To address this limitation, this paper proposes a novel spatial-frequency cued GFANC (SF-GFANC) method that exploits both three-dimensional (3D) spatial and frequency information of the noise source. Specifically, a multi-task convolutional recurrent neural network (CRNN) is designed to estimate the source distance, elevation angle, and azimuth angle as spatial cues, while predicting the combination weights of sub control filters as frequency cues. These spatial-frequency cues jointly guide the generation of the appropriate control filter. In addition, a theoretical analysis of the optimal control filter in reverberant environments is presented, highlighting the importance of 3D spatially conditioned control filter design. Evaluations using both simulated and measured acoustic paths demonstrate that the CRNN is robust to unseen acoustic environments and noise types. Furthermore, the results confirm that SF-GFANC outperforms representative ANC algorithms when handling noise sources across diverse 3D locations and frequency characteristics in reverberant environments.\fntext[]{The code will be available at \href{https://github.com/Wang-Boxiang/Spatial-Frequency-Generative-Fixed-Filter-Active-Noise-Control}{https://github.com/Wang-Boxiang/Spatial-Frequency-Generative-Fixed-Filter-Active-Noise-Control}}

\end{abstract}



\begin{highlights}
\item A spatial–frequency cued GFANC method is proposed to address the lack of spatial awareness in the vanilla GFANC method.

\item Theoretical analysis establishes the role of 3D spatial conditioning in optimal control filter design under reverberation.

\item A multi-task CRNN jointly learns 3D spatial and frequency cues, enabling accurate control filter generation in reverberant environments.

\item Results on simulated and measured acoustic paths confirm superior robustness and performance across diverse noise locations and spectra.
\end{highlights}


\end{frontmatter}

\section{Introduction}
Active noise control (ANC) is an advanced technique that can effectively attenuate low-frequency noise through the principle of sound destructive interference, in which a secondary source generates anti-noise with equal amplitude and opposite phase to the unwanted noise~\cite{elliott1993active}. Compared with passive methods that rely on bulky barriers, ANC provides a more compact and effective solution and has therefore been applied in various applications~\cite{cheer2019application,zhang2018active}.  

Traditional ANC systems often employ adaptive algorithms, such as the filtered-reference least mean squares (FxLMS), to update the control filter in real time \cite{dong2020distributed,liu2024study}. However, these algorithms often involve slow convergence speed and are at risk of divergence \cite{zhang2021deep,xie2024cognitive}. Furthermore, adaptive algorithms require the placement of an error microphone at the target location, which imposes physical constraints \cite{ji2025preventing,li2023augmented}. Recently, deep learning (DL) has emerged as a promising alternative to improve performance for both active noise and vibration control \cite{fareedha2025dynamic,oh2026head,zhao2025advances,yang2024data}, demonstrating strong capabilities in complex environment mapping and nonlinear modeling. To address these limitations, the DL-based selective fixed-filter ANC (SFANC) method has been proposed to select the most suitable pre-trained control filter for various noise types \cite{shi2022selective}. By eliminating the need for error microphone placement, this method improves practicality while ensuring high stability and fast response. To further enhance noise reduction performance while avoiding extensive control filter training, the DL-based generative fixed-filter ANC (GFANC) method has been proposed to generate a more suitable control filter \cite{luo2023deep,luo2025mssp}. However, both SFANC and GFANC primarily focus on the frequency characteristics of the noise source and overlook spatial information, which has also been proven to be critical for the performance of ANC systems \cite{liebich2018direction,ho2021time}.

To date, several researchers have incorporated the influence of spatial variation of the noise source into ANC systems. Some studies have aimed to improve noise cancellation for sources arriving from different directions by enhancing the causality of ANC systems \cite{zhang2014causality}. Others have focused on achieving spatial selectivity, in which unwanted noise is suppressed while the desired sound is preserved \cite{zhang2023time,xiao2023spatially}. However, these techniques rely on real-time control filter updates, which often suffer from slow response and potential stability issues. As an alternative, Toyooka et al. proposed a method that selects control filters corresponding to different noise source locations \cite{toyooka2025active}. In contrast, Su et al. and Luo et al. considered both the frequency and direction-of-arrival (DoA) information of the noise source for control filter selection, utilizing either a traditional signal processing technique \cite{su2024spatial} or a DL-based approach \cite{luo2025doa}. Nevertheless, these methods are formulated based on free-field assumptions, whereas in real-world scenarios, noise sources are often encountered in reverberant environments such as offices or living rooms \cite{aboutiman2025subjective}. 

Although the recent study \cite{wangdirectional} explored a DL-based approach to mitigate indoor noise sources with different DoAs, it failed to account for the source frequency and distance, thereby limiting its practical applicability. In particular, the estimation of noise source distance has not been thoroughly investigated in existing ANC literature, which remains a challenging task. Moreover, the theoretical foundation for employing appropriate control filters for noise sources at different three-dimensional (3D) locations in reverberant environments has not been systematically established. Therefore, it is necessary to conduct a comprehensive analysis of the relationship between the optimal control filter and the 3D source location, and to develop a unified ANC framework that can effectively suppress noises across different 3D positions and frequency characteristics.

To address these limitations, this paper proposes a novel spatial-frequency cued GFANC (SF-GFANC) method that jointly exploits the 3D spatial and frequency information of the noise source. In this framework, given multi-channel reference signals as input, a multi-task CRNN is employed to estimate the source distance, elevation angle, and azimuth angle, enabling adaptation to different 3D source locations. Meanwhile, it predicts the combination weights of sub control filters, allowing the system to accommodate noise sources with diverse frequency content. By integrating both spatial and frequency cues, the appropriate control filter is generated at the frame level and applied on a sample-by-sample basis to achieve delayless noise control in reverberant environments. It should be noted that the proposed framework is designed for single-source scenarios, which is consistent with the system formulation considered in this work. The main contributions of this paper are summarized as follows:
\begin{itemize}
    \item Unlike conventional GFANC methods, the proposed SF-GFANC framework introduces a joint spatial-frequency formulation. By integrating 3D spatial cues with frequency characteristics, it accurately generates appropriate control filters for complex reverberant environments.
    \item A theoretical analysis is presented to establish the relationship between the optimal control filter and the 3D spatial coordinates of the noise source in reverberant environments, providing a principled foundation for spatially conditioned control filter design.
    \item A multi-task CRNN is proposed to jointly estimate 3D spatial and frequency cues. This unified framework enables accurate control filter generation under varying acoustic conditions without requiring separate models.
    \item Extensive experiments on both simulated and measured acoustic paths demonstrate that the proposed SF-GFANC method achieves improved noise reduction performance compared to representative ANC baselines.
\end{itemize}

The remainder of this paper is organized as follows. Section~\ref{Preliminaries} introduces the fundamentals of DL-based 3D sound source localization and multi-reference ANC systems. Section~\ref{theory} theoretically analyzes the importance of employing the appropriate control filter for noise sources located at different 3D locations in reverberant environments. Section~\ref{proposed method} describes the details of the proposed SF-GFANC method. Section~\ref{simulations} evaluates the performance of the proposed algorithm through numerical simulations. Finally, Section~\ref{conclusion} concludes the whole paper.

\section{Preliminaries}
\label{Preliminaries}
This section outlines the basics of DL-based 3D sound source localization and multi-reference ANC systems. The limitations of traditional 3D localization approaches and adaptive multi-reference ANC algorithms are also discussed.

\subsection{DL-based 3D Sound Source Localization}
To determine a source position in 3D space, DoA estimation needs to be performed together with source distance estimation. Traditional DoA estimation algorithms, such as steered-response power (SRP), have achieved notable progress over the years. However, these methods are typically formulated under the assumption of free-field sound propagation and often perform poorly under reverberant conditions with background noise \cite{li2023doa}. In contrast, source distance estimation is considered as a more challenging task. Conventional approaches, such as those based on the direct-to-reverberant ratio (DRR), typically require prior knowledge of the acoustic environment or sound source direction \cite{krause2023binaural}.


Unlike traditional methods, data-driven DL-based 3D sound source localization methods mitigate the dependency on explicit propagation models and environmental priors, making them well-suited for complex real-world acoustic scenarios \cite{feng2025eliminating,chakrabarty2019multi}. In reverberant environments, variations in source location, microphone placement, and room acoustics lead to each microphone receiving the source signal convolved with a distinct room impulse response (RIR). The resulting interchannel differences in delay and amplitude encode spatial information about the source location. Given a large amount of representative training data, a deep neural network (DNN) can automatically identify the relationship between the multichannel signal $\bf{r}$ and the 3D sound source location. The DoA and distance can be estimated in a unified framework as
\begin{equation}
\setlength{\abovedisplayskip}{2pt}
\setlength{\belowdisplayskip}{2pt}
(l,\phi,\theta) = {\rm{DNN}}({\bf{r}}),
\end{equation}
where $l$, $\phi$ and $\theta$ denote the distance, elevation angle, and azimuth angle of the source relative to the microphone array. In particular, CRNNs have been widely applied to 3D sound source localization tasks due to their capability to extract spatial features and capture temporal dependencies \cite{perotin2019crnn,neri2024speaker}. Accordingly, the proposed SF-GFANC framework employs a multi-task CRNN to jointly estimate the 3D location along with the spectral content of the noise source for accurate control filter generation in reverberant environments.
\begin{figure}[!t]
\centering
\includegraphics[width=4in]{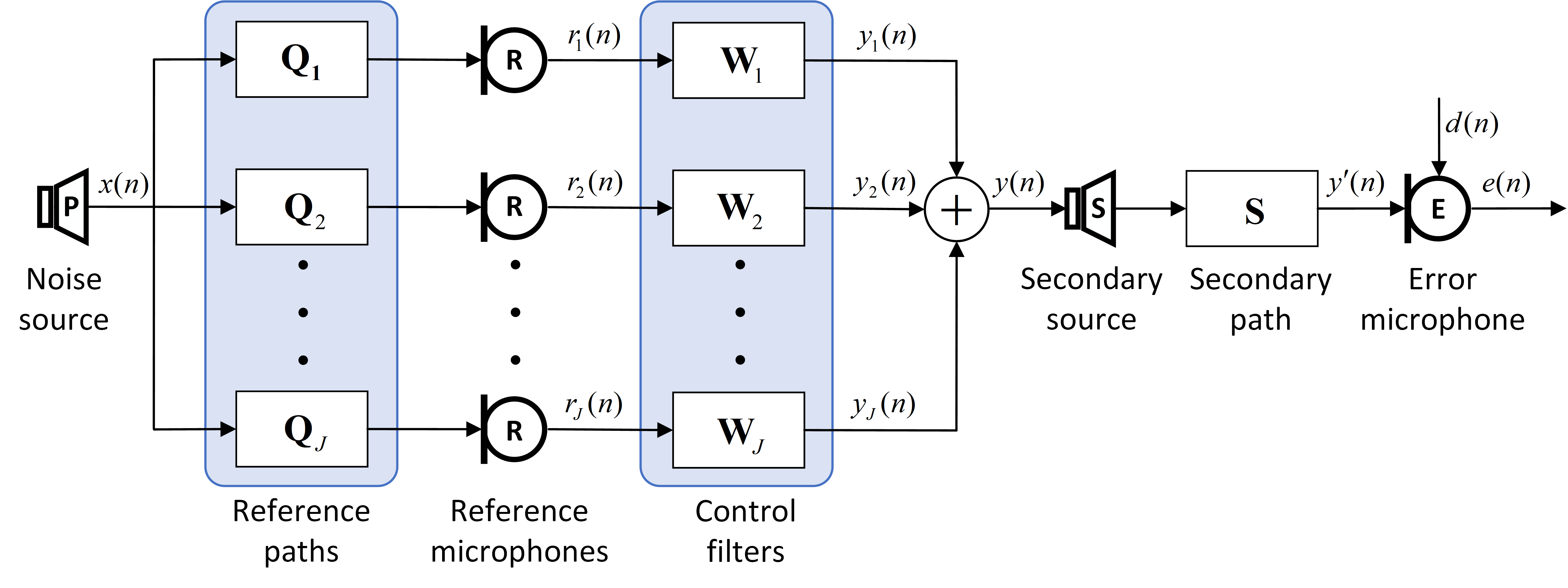}\vspace*{-0.4cm}
\caption{Block diagram of a multi-reference active noise control system with $J$ reference microphones, one secondary source and one error microphone.}
\label{fig_1}
\end{figure}
\subsection{Multi-Reference Active Noise Control Systems}
\label{section:adaptive}
In addition to sound source localization, the multichannel signals can also serve as reference inputs for the ANC system. Fig.~\ref{fig_1} illustrates a multi-reference ANC system with $J$ reference microphones, one secondary source and one error microphone. The control filter vector is given by
\begin{equation}
\setlength{\abovedisplayskip}{2pt}
\setlength{\belowdisplayskip}{2pt}
{\bf{w}}(n) = \left[ {{\bf{w}}_1(n),{\bf{w}}_2(n),\ldots,{\bf{w}}_J(n)} \right] \in \mathbb{R} {^{1 \times JL_{\rm c}}},
\end{equation}
where ${{\mathbf{w}}_j}(n) = {[{w_{j,1}}(n),{w_{j,2}}(n),\cdots,{w_{j,L_{\rm c}}}(n)]} \in \mathbb{R} {^{1 \times L_{\rm c}}}$ is the $j$-th $(j = 1,2,\ldots,J)$ control filter vector, $\mathbb{R} {^{1 \times L_{\rm c}}}$ denotes a $1 \times L_{\rm c}$ vector of real values, and $L_{\rm c}$ is the control filter length. The reference signal vector is given by
\begin{equation}
\setlength{\abovedisplayskip}{2pt}
\setlength{\belowdisplayskip}{2pt}
{\bf{r}}(n) = \left[ {{\bf{r}}_1(n),{\bf{r}}_2(n),\ldots,{\bf{r}}_J(n)} \right]^{\rm T} \in \mathbb{R} {^{JL_{\rm c} \times 1}},
\end{equation}
where ${{\mathbf{r}}_j}(n) = {[{r_j}(n),{r_j}(n - 1),\cdots,{r_j}(n - L_{\rm c} + 1)]} \in \mathbb{R} {^{1 \times L_{\rm c}}}$ is the $j$-th reference signal vector, and ${( \cdot )^ {\rm T} }$ denotes the transpose operator.

The control signal utilized to drive the secondary source is expressed as
\begin{equation}
\setlength{\abovedisplayskip}{2pt}
\setlength{\belowdisplayskip}{2pt}
y(n) = {{\bf{w}}}(n){\bf{r}}(n).
\end{equation}

The resulting error signal can be expressed as
\begin{equation}
\setlength{\abovedisplayskip}{2pt}
\setlength{\belowdisplayskip}{2pt}
\begin{aligned}
e(n) &= d(n) - y'(n) = d(n) - s(n) * y(n),
\label{eq:1}
\end{aligned}
\end{equation}
where $d(n)$ is the disturbance at the error microphone, $y'(n)$ is the anti-noise, $s(n)$ is the secondary path impulse response, and $*$ is the convolution operator.

According to the FxLMS algorithm, the control filter vector is updated by
\begin{equation}
\setlength{\abovedisplayskip}{2pt}
\setlength{\belowdisplayskip}{2pt}
{\bf{w}}(n + 1) = {\bf{w}}(n) + \mu {\bf{r'}}(n)e(n),
\label{eq:fxlms}
\end{equation}
where $\mu$ is the step size, and ${{{\mathbf{r}}}'}(n)$ is the filtered reference signal vector obtained through the estimated secondary path $\hat s(n)$ as 
\begin{equation}
\setlength{\abovedisplayskip}{2pt}
\setlength{\belowdisplayskip}{2pt}
{\bf{r'}}(n) = \hat s(n) * {\bf{r}}(n)^{\rm T}\in \mathbb{R} {^{1 \times JL_{\rm c}}}.
\label{eq:2}
\end{equation}

However, the adaptive FxLMS algorithm requires the placement of an error microphone, suffers from slow convergence and may lead to noise divergence if the step size is not properly chosen. To alleviate these problems, the proposed SF-GFANC method generates the appropriate control filter based on both the 3D spatial and frequency information of the noise source.

\section{Optimal Control Filter for Multi-Reference ANC Systems in Reverberant Environments}
\label{theory}
Although prior studies have established that the optimal control filter depends on the acoustic paths~\cite{elliott1993active,zhang2018active} and the noise spectrum~\cite{shi2022selective}, its dependence on the source location is typically expressed through the underlying path transfer functions. In reverberant environments, an explicit formulation that relates the optimal control filter directly to the noise source's 3D location has not been systematically developed. In this section, it is shown that the conventional path-based expression can be reformulated into an explicit dependence on 3D spatial parameters, including distance, elevation angle, and azimuth angle, thereby providing a principled basis for 3D spatially conditioned control filter design.

The analysis assumes that the locations of the reference microphones, secondary source, and error microphone are fixed. The error signal in the frequency domain is represented as
\begin{equation}
\setlength{\abovedisplayskip}{2pt}
\setlength{\belowdisplayskip}{2pt}
E(\omega )=D(\omega )-S(\omega )\bf{W}(\omega )\bf{R}(\omega ),
\end{equation}
where $D(\omega )$ is the frequency response of the disturbance, $S(\omega )$ is the secondary path, $\mathbf{W}(\omega )\in \mathbb{C} {^{1 \times J}}$ is the control filter, with $ \mathbb{C} {^{1 \times J}}$ representing a $1 \times J$ vector of complex values, and $\mathbf{R}(\omega )\in \mathbb{C} {^{J \times 1}}$ is the reference signal.

To find the optimal control filter, the cost function is the power spectral density (PSD) of the error signal, given by
\begin{equation}
\setlength{\abovedisplayskip}{2pt}
\setlength{\belowdisplayskip}{2pt}
\begin{aligned}
J(\omega) &= {{C}_{dd}}(\omega )-{{\bf{C}}_{{\bf{r}}d}}(\omega )\mathbf{W}^{\mathrm{H}}(\omega)S(\omega )^ * -S(\omega )\mathbf{W}(\omega ){{\bf{C}}^{\mathrm{H}}_{{\bf{r}}d}}(\omega ) \\
& + S(\omega )\mathbf{W}(\omega ){{\bf{C}}_{{\bf{rr}}}}(\omega )\mathbf{W}^{\mathrm{H}}(\omega)S(\omega )^ *,
\end{aligned}
\end{equation}
where ${( \cdot )^ {\rm H} }$ is the Hermitian transpose operator, ${( \cdot )^ * }$ is the conjugation operator, ${{C}_{dd}}(\omega )$ is the PSD of the disturbance, ${{\bf{C}}_{{\bf{r}}d}}(\omega )$ is the cross spectral density between the reference signal and the disturbance, and ${{\bf{C}}_{{\bf{rr}}}}(\omega )$ is the PSD of the reference signal, defined as
\begin{equation}
\setlength{\abovedisplayskip}{2pt}
\setlength{\belowdisplayskip}{2pt}
{{C}_{dd}}(\omega ) = \mathbb{E}[D(\omega )D^{\mathrm{H}}(\omega )],
\end{equation}
\begin{equation}
\setlength{\abovedisplayskip}{2pt}
\setlength{\belowdisplayskip}{2pt}
{{\bf{C}}_{{\bf{r}}d}}(\omega ) = \mathbb{E}[D(\omega )\mathbf{R}^{\mathrm{H}}(\omega )]\in \mathbb{C} {^{1 \times J}},
\end{equation}
\vspace*{-0.7cm}
\begin{equation}
\setlength{\abovedisplayskip}{2pt}
\setlength{\belowdisplayskip}{2pt}
{{\mathbf{C}}_{{\bf{rr}}}}(\omega ) = \mathbb{E}[\mathbf{R}(\omega )\mathbf{R}^{\mathrm{H}}(\omega )]\in \mathbb{C} {^{J \times J}},
\end{equation}
where $\mathbb{E}[ \cdot ]$ denotes the expectation operator and the optimal control filter is obtained when the cost function reaches its minimum value. The gradient of $J(\omega)$ with respect to the control filter $\bf{W}(\omega )$ is expressed as
\begin{equation}
\setlength{\abovedisplayskip}{2pt}
\setlength{\belowdisplayskip}{2pt}
\nabla J(\omega) =  - 2[S(\omega )^ *{{\bf{C}}_{{\bf{r}}d}}(\omega ) - S(\omega )^ *S(\omega )\bf{W}(\omega ){{\bf{C}}_{{\bf{rr}}}}(\omega )].
\end{equation}

When $\nabla J(\omega)$ is set to zero, the optimal control filter can be obtained as
\begin{equation}
\setlength{\abovedisplayskip}{2pt}
\setlength{\belowdisplayskip}{2pt}
{{\bf{W}}^o}(\omega ) = S^{ - 1}(\omega ){{\bf{C}}_{{\bf{r}}d}}(\omega ){{\bf{C}}^{ - 1}_{{\bf{rr}}}}(\omega ) \in \mathbb{C} {^{1 \times J}}.
\label{eq:3}
\end{equation}

Since the disturbance $D(\omega )$ and the reference signal $\mathbf{R}(\omega )$ are generated by passing the noise signal $X(\omega)$ through the primary path $P(\omega )$ and the reference path ${\mathbf{Q}}(\omega )\in \mathbb{C} {^{J \times 1}}$ respectively, the cross spectral density ${{\bf{C}}_{{\bf{r}}d}}(\omega )$ and the PSD ${{\bf{C}}_{{\bf{rr}}}}(\omega )$ can be expressed as \cite{shi2020virtualjasa}
\begin{equation}
\setlength{\abovedisplayskip}{2pt}
\setlength{\belowdisplayskip}{2pt}
{{\bf{C}}_{{\bf{r}}d}}(\omega ) = P(\omega ){{C}_{xx}}(\omega ){\bf{Q}}^{\mathrm{H}}(\omega )\in \mathbb{C} {^{1 \times J}},
\label{eq:4}
\end{equation}
\begin{equation}
\setlength{\abovedisplayskip}{2pt}
\setlength{\belowdisplayskip}{2pt}
{{\bf{C}}_{{\bf{rr}}}}(\omega ) = {\bf{Q}}(\omega ){{C}_{xx}}(\omega ){\bf{Q}}^{\mathrm{H}}(\omega )\in \mathbb{C} {^{J \times J}},
\label{eq:5}
\end{equation}
where ${C_{xx}}(\omega)$ is the PSD of the noise signal.

Substituting Eqs.~\eqref{eq:4} and~\eqref{eq:5} into Eq.~\eqref{eq:3}, the frequency response of the optimal control filter can be rewritten as
\begin{equation}
\setlength{\abovedisplayskip}{2pt}
\setlength{\belowdisplayskip}{2pt}
{{\bf{W}}^o}(\omega ) = \frac{{P(\omega )}}{{S(\omega )}}{[{\bf{Q}}^{\rm H}(\omega ){{\bf{Q}}}(\omega )]^{ - 1}}{{\bf{Q}}^{\rm H} }(\omega )\in \mathbb{C} {^{1 \times J}}.
\label{eq:10}
\end{equation}

Given that most indoor reverberant environments can be approximately modeled as rectangular enclosures with non-rigid boundaries, the image method provides a widely used approach for room acoustics analysis \cite{allen1979image}, where the RIR between the noise source and the error microphone is expressed as
\begin{equation}
\setlength{\abovedisplayskip}{2pt}
\setlength{\belowdisplayskip}{2pt}
p(n) = \sum\limits_{k = 1}^K {{\alpha _k}\frac{{\delta (n - {n_k})}}{{4\pi {\rho _k}}}},
\label{eq:6}
\end{equation}
where ${n_k} = \left\lfloor {\frac{{{\rho _k}}}{{c_{\rm{a}}T_{\rm s}}}} \right\rfloor$. ${{\alpha _k}}$ is the gain for the $k$-th propagation path, determined by the reflection coefficients of the six walls. ${{\rho _k}}$ is the length between the image source and the error microphone for the $k$-th propagation path, which depends on the room geometry and the locations of the noise source and the error microphone. $K$ is the total number of propagation paths between the noise and the error microphone, $c_{\rm{a}}$ is the sound speed in air, $T_{\rm s}$ is the sampling period, $\left\lfloor \cdot \right\rfloor$ is the floor function, and $\delta ( \cdot )$ is the Dirac delta function.

Similarly, the RIR between the noise source and the $j$-th reference microphone is stated as 
\begin{equation}
\setlength{\abovedisplayskip}{2pt}
\setlength{\belowdisplayskip}{2pt}
{q_j}(n) = \sum\limits_{m = 1}^{{M_j}} {{\beta _{j,m}}\frac{{\delta (n - {n_{j,m}})}}{{4\pi {\upsilon _{j,m}}}}}, \quad j = 1,2,\ldots,J,
\label{eq:7}
\end{equation}
where ${n_{j,m}} = \left\lfloor {\frac{{{\upsilon_{j,m}}}}{{c_{\rm{a}}T_{\rm s}}}} \right\rfloor$. ${{\beta _{j,m}}}$ is the gain for the $m$-th propagation path for the $j$-th reference microphone. ${{\upsilon_{j,m}}}$ is the length between the image source and the $j$-th reference microphone for the $m$-th propagation path. $M_j$ is the total number of propagation paths between the noise and the $j$-th reference microphone.

The discrete-time Fourier transform of Eqs.~\eqref{eq:6} and~\eqref{eq:7} can be written as 
\begin{equation}
\setlength{\abovedisplayskip}{2pt}
\setlength{\belowdisplayskip}{2pt}
P(\omega ) = \sum\limits_{k = 1}^{K} {{\alpha _k}\frac{{{e^{ - i\frac{\omega }{c_{\rm{a}}}{\rho _k}}}}}{{4\pi {\rho _k}}}},
\label{eq:8}
\end{equation}
\begin{equation}
\setlength{\abovedisplayskip}{2pt}
\setlength{\belowdisplayskip}{2pt}
Q_j(\omega ) = \sum\limits_{m = 1}^{M_j} {{\beta _{j,m}}\frac{{{e^{ - i\frac{\omega }{c_{\rm{a}}}{\upsilon_{j,m}}}}}}{{4\pi {\upsilon_{j,m}}}}}, \quad j = 1,2,\ldots,J,
\label{eq:9}
\end{equation}
where $i = \sqrt { - 1}$.

Substituting Eqs.~\eqref{eq:8} and~\eqref{eq:9} into Eq.~\eqref{eq:10}, the frequency response of the $j$-th optimal control filter $W_j^o(\omega )$ can be rewritten as
\begin{equation}
\setlength{\abovedisplayskip}{2pt}
\setlength{\belowdisplayskip}{2pt}
\begin{aligned}
W_j^o(\omega ) &= \frac{{P(\omega )}}{{S(\omega )}}\frac{{Q_j^ * (\omega )}}{{\sum\limits_{u = 1}^J {{{\left| {{Q_u}(\omega )} \right|}^2}} }} \\
&= \frac{{\sum\limits_{k = 1}^{K} {{\alpha _k}\frac{{{e^{ - i\frac{\omega }{c_{\rm{a}}}{\rho _k}}}}}{{4\pi {\rho _k}}}} }}{{S(\omega )}} \times \frac{{\sum\limits_{m = 1}^{{M_j}} {{\beta _{j,m}}\frac{{{e^{i\frac{\omega }{c_{\rm{a}}}{\upsilon_{j,m}}}}}}{{4\pi {\upsilon_{j,m}}}}} }}{{\sum\limits_{u = 1}^J {\sum\limits_{m = 1}^{{M_u}} {\sum\limits_{{m'} = 1}^{{M_u}} {{\beta _{u,m}}{\beta _{u,{m'}}}\frac{{{e^{ - i\frac{\omega }{c_{\rm{a}}}({\upsilon_{u,m}} - {\upsilon_{u,{m'}}})}}}}{{{{(4\pi )}^2}{\upsilon_{u,m}}{\upsilon_{u,{m'}}}}}} } } }}.
\end{aligned}
\label{eq:11}
\end{equation}




From Eq.~\eqref{eq:11}, it can be seen that, for a multi-reference ANC system operating in a specific reverberant environment, the location of the noise source plays a critical role. When the locations of the reference microphones and the error microphone are fixed, the noise source location determines the propagation path lengths for both the primary path ($\rho_k$) and the reference paths ($\upsilon_{j,m}$, $\upsilon_{u,m}$ and $\upsilon_{u,m'}$), therefore jointly influencing the frequency response of the optimal control filter. In other words, the optimal control filter depends on the distance, elevation angle, and azimuth angle of the noise source relative to the reference microphones. Therefore, if the noise source location varies over time in a reverberant environment, using a fixed pre-trained control filter may result in degraded noise reduction performance. It is thus essential to employ a control filter that is suitable for the noise source location to achieve effective noise reduction.

\begin{figure}[!t]
\centering
\includegraphics[width=5.5in]{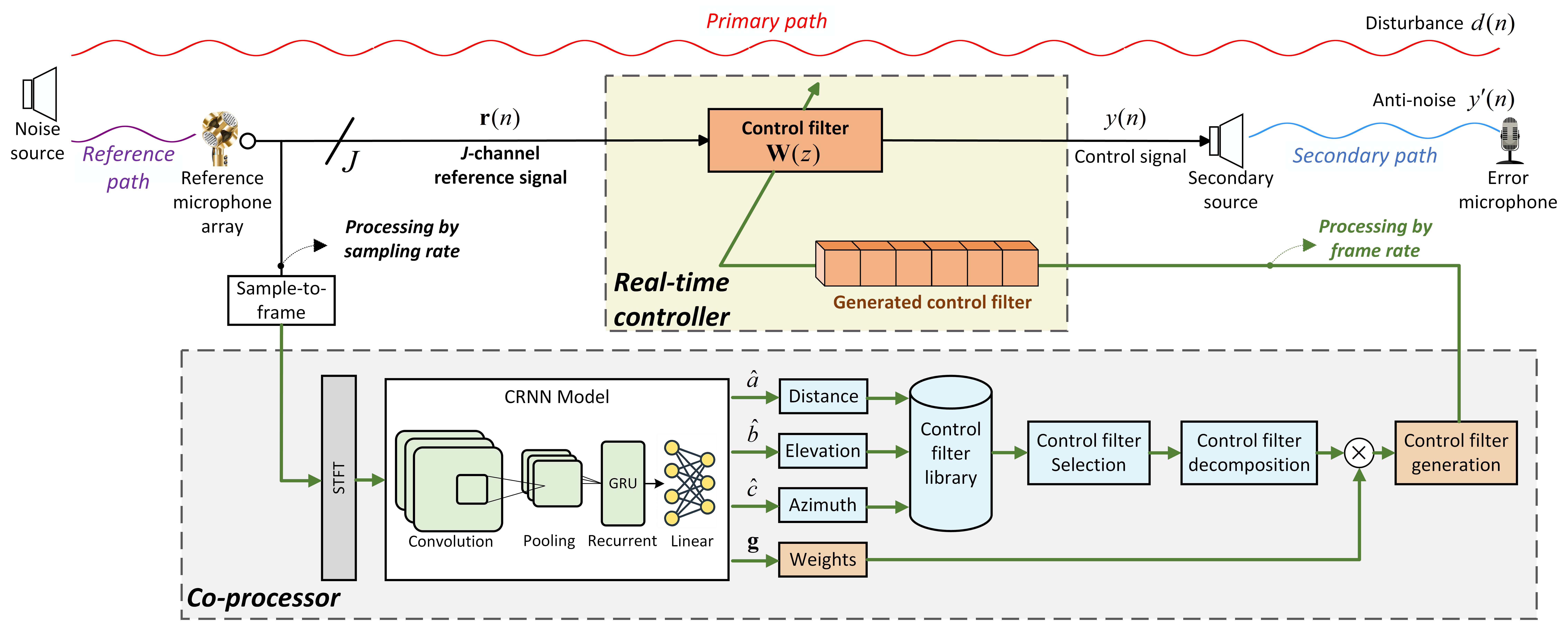}\vspace*{-0.4cm}
\caption{Block diagram of the proposed SF-GFANC method. A co-processor executes the CRNN to perform distance, elevation angle, azimuth angle classifications, and combination weights regression, followed by control filter generation for each noise frame. In parallel, real-time noise control is performed at the sampling rate. The error microphone is used solely to monitor noise reduction performance and is not involved in the control filter update process.}
\label{fig_2}
\end{figure}

\section{The Proposed SF-GFANC Method}
\label{proposed method}
This section details the proposed SF-GFANC method. To translate the theoretical spatial dependency established in Section 3 into a practical implementation, the proposed framework approximates this dependency by explicitly constructing a spatially sampled control filter library and employing a multi-task CRNN for the estimation of spatial cues. As illustrated in Fig.~\ref{fig_2}, the co-processor and the real-time controller operate in parallel at different rates to enable delayless noise control. Specifically, the generated control filter incorporates both spatial and frequency cues of the noise source, enabling effective attenuation of noise sources with various 3D locations and frequency characteristics in reverberant environments.

\subsection{Pre-trained Control Filter Library}
Prior to the online execution of the SF-GFANC method, a control filter library must be pre-trained to accommodate noise sources located at various 3D locations. Specifically, the noise sources are placed on spherical grids surrounding the reference microphone array center, defined by $A$ discrete distances, $B$ elevation angles, and $C$ azimuth angles. The values of $A$, $B$, and $C$ represent a trade-off between spatial resolution and storage cost and can be adjusted according to system requirements. At each grid point, a control filter $\left[ {{{\bf{w}}^{a,b,c}}} \right]_{a = 1,b = 1,c = 1}^{A,B,C} \in \mathbb{R} {^{1 \times JL_{\rm c}}}$ is pre-trained for a broadband white noise using the FxLMS algorithm described in Section~\ref{section:adaptive}. The resulting control filters are stored in a library for subsequent deployment during online noise control.

\vspace*{-0.3cm}
\subsection{CRNN for Spatial-Frequency Cues Estimation}
Subsequently, a CRNN is trained using a multi-task learning strategy to perform distance, elevation angle, and azimuth angle classifications, along with combination weights regression. These spatial-frequency cues jointly guide the generation of the appropriate control filter.

\vspace*{-0.3cm}
\subsubsection{CRNN Architecture}
\begin{figure}[!t]
\centering
\includegraphics[width=5.5in]{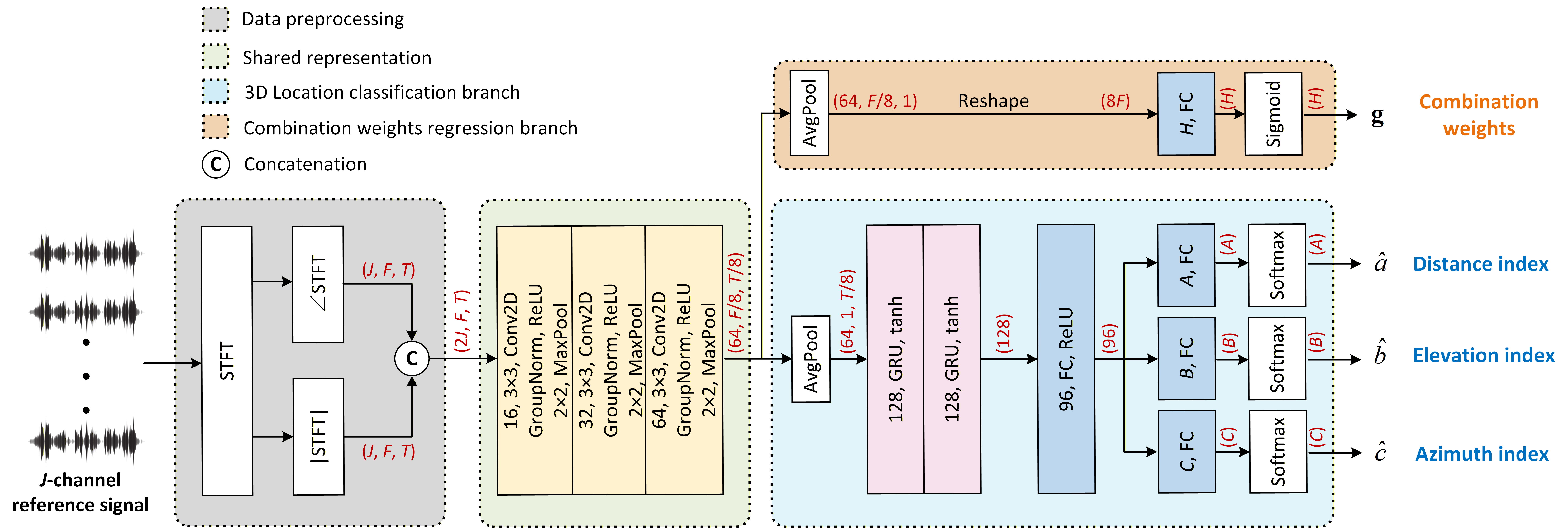}\vspace*{-0.4cm}
\caption{Proposed CRNN architecture for 3D location classification and combination weights regression.}
\label{fig_3}
\end{figure}

The architecture of the proposed CRNN is illustrated in Fig.~\ref{fig_3}. Instead of employing an explicit feature extraction pipeline, the short-time Fourier transform (STFT) is applied for data preprocessing, allowing the convolutional layers to learn the spatial and frequency characteristics directly from data. Specifically, the STFT is computed for each frame of the $J$-channel reference signal ${{\bf{r}}} \in \mathbb{R} {^{J \times L_{\rm r}}}$, denoted as $\text{STFT}({\bf{r}}) \in \mathbb{C} {^{J \times F \times T}}$, where $L_{\rm r}$ is the number of samples, $F$ is the number of frequency bins and $T$ is the number of time frames. From the STFT output, both the magnitude map $\left| {{\rm{STFT}}({\bf{r}})} \right| \in \mathbb{R} {^{J \times F \times T}}$ and phase map $\angle {\rm{STFT}}({\bf{r}}) \in \mathbb{R} {^{J \times F \times T}}$ are computed. These are then concatenated along the channel dimension to form a $2 J \times F \times T$ tensor and fed into the CRNN. 

To extract spatial and spectral features, the pre-processed input is passed through three convolutional blocks, each comprising a two-dimensional (2D) convolutional layer, group normalization, ReLU activation, and max-pooling along both frequency and time axes. The resulting feature maps are then fed into two task-specific branches: one for combination weights regression and the other for 3D location classification. In the combination weights regression branch, the output of the convolutional blocks is first averaged over time and reshaped. A fully connected (FC) layer with sigmoid activation is then applied to estimate the combination weights of the sub control filters as frequency cues. In the 3D location classification branch, average pooling is first applied along the frequency axis to reduce dimensionality. The resulting feature maps are then passed through two recurrent blocks, each consisting of a gated recurrent unit (GRU) layer with tanh activation, to capture temporal dependencies. A subsequent FC layer compresses the recurrent output, which is then fed into three parallel FC layers that classify the distance, elevation angle, and azimuth angle of the noise source. Final predictions for each class are obtained using softmax activation functions as spatial cues.

\subsubsection{CRNN Training Objective}
As previously discussed, the CRNN is trained using a multi-task learning strategy that jointly performs distance, elevation angle, and azimuth angle classifications alongside combination weights regression. Specifically, the cross-entropy loss is used for the classification tasks \cite{shi2022selective}, while the regression task is optimized using the mean squared error (MSE) loss \cite{luo2023delayless}. The overall training objective is defined as the sum of these losses. This joint loss allows the CRNN to learn shared representations while effectively managing multiple objectives. Moreover, it provides a more efficient alternative to training separate models for each task, which would be significantly more time-consuming and computationally demanding.

\subsection{Control Filter Selection}
\label{sec:selection}
After training the CRNN, given the input tensor $\mathbf{R}\in \mathbb{R} {^{2J \times F \times T}}$, the CRNN outputs the predicted probabilities for each distance class, elevation angle class, azimuth angle class and the soft combination weights of the sub control filters, expressed as
\begin{equation}
\setlength{\abovedisplayskip}{2pt}
\setlength{\belowdisplayskip}{2pt}
({{{\bf{\hat p}}}_{{\rm{dist}}}},{{{\bf{\hat p}}}_{{\rm{elev}}}},{{{\bf{\hat p}}}_{{\rm{azim}}}},{\bf{\hat g}}) = {\rm{CRNN}}({\bf{R}};{\Theta ^*}),
\end{equation}
where
\begin{equation}
\setlength{\abovedisplayskip}{2pt}
\setlength{\belowdisplayskip}{2pt}
\left\{ \begin{array}{l}
{{{\bf{\hat p}}}_{{\rm{dist}}}} = [{{\hat p}_{{\rm{dist}},1}},{{\hat p}_{{\rm{dist}},2}},\ldots,{{\hat p}_{{\rm{dist}},A}}]\in \mathbb{R} {^{1 \times A}},\\
{{{\bf{\hat p}}}_{{\rm{elev}}}} = [{{\hat p}_{{\rm{elev}},1}}, {{\hat p}_{{\rm{elev}},2}}, \ldots ,{{\hat p}_{{\rm{elev}},B}}]\in \mathbb{R} {^{1 \times B}},\\
{{{\bf{\hat p}}}_{{\rm{azim}}}} = [{{\hat p}_{{\rm{azim}},1}}, {{\hat p}_{{\rm{azim}},2}}, \ldots ,{{\hat p}_{{\rm{azim}},C}}]\in \mathbb{R} {^{1 \times C}},\\
{\bf{\hat g}} = \left[ {{\hat g_1}, {\hat g_2},\ldots ,{\hat g_H}} \right]\in \mathbb{R} {^{1 \times H}},
\end{array} \right.
\end{equation}
where ${\Theta ^*}$ denotes the trained parameters of the CRNN; ${{{\bf{\hat p}}}_{{\rm{dist}}}}$, ${{{\bf{\hat p}}}_{{\rm{elev}}}}$, and ${{{\bf{\hat p}}}_{{\rm{azim}}}}$ represent the predicted probability distributions over the $A$ distance classes, $B$ elevation angle classes, and $C$ azimuth angle classes, respectively; and ${\bf{\hat g}}$ is the soft combination weight vector of the $H$ sub control filters.

The estimated distance index $\hat a$, elevation angle index $\hat b$ and azimuth angle index $\hat c$ can be obtained as
\begin{equation}
\setlength{\abovedisplayskip}{2pt}
\setlength{\belowdisplayskip}{2pt}
\left\{ \begin{array}{l}
\hat a = \mathop {\arg \max }\limits_{{i_{\rm a}} \in \{ 1,2,\ldots,A\} } {{\hat p}_{{\rm{dist}},{i_{\rm a}}}},\\
\hat b = \mathop {\arg \max }\limits_{{i_{\rm b}} \in \{ 1,2,\ldots,B\} } {{\hat p}_{{\rm{elev}},{i_{\rm b}}}},\\
\hat c = \mathop {\arg \max }\limits_{{i_{\rm c}} \in \{ 1,2,\ldots,C\} } {{\hat p}_{{\rm{azim}},{i_{\rm c}}}},
\end{array} \right.
\end{equation}
where ${{\hat p}_{{\rm{dist}},{i_{\rm a}}}}$, ${{\hat p}_{{\rm{elev}},{i_{\rm b}}}}$, and ${{\hat p}_{{\rm{azim}},{i_{\rm c}}}}$ denote the predicted probabilities of the $i_{\rm a}$-th distance, $i_{\rm b}$-th elevation angle, and $i_{\rm c}$-th azimuth angle classes, respectively.

Based on the predicted class labels for distance, elevation angle and azimuth angle, the corresponding control filter ${{\bf{w}}^{\hat a,\hat b,\hat c}} \in \mathbb{R} {^{1 \times JL_{\rm c}}}$ is retrieved from the pre-trained library. This selection process leverages the 3D spatial information of the noise source to ensure that the selected control filter matches its location.

\vspace*{-0.3cm}
\subsection{Control Filter Generation}
\vspace*{-0.1cm}
\label{sec:generation}
Following that, the selected control filter ${{\bf{w}}^{\hat a,\hat b,\hat c}} \in \mathbb{R}^{1 \times JL_{\rm c}}$ is decomposed into $H$ sub control filters based on the theory of filter perfect reconstruction \cite{luo2023delayless}. In the frequency domain, its discrete Fourier transform (DFT) can be interpreted as the sum of the DFTs of these $H$ sub control filters, expressed as
\begin{equation}
\setlength{\abovedisplayskip}{2pt}
\setlength{\belowdisplayskip}{2pt}
{{\bf{W}}^{\hat a,\hat b,\hat c}} = \sum\limits_{h = 1}^H {{\bf{C}}_h^{\hat a,\hat b,\hat c}}\in \mathbb{C}^{1 \times J},
\end{equation}
where ${{\bf{C}}_h^{\hat a,\hat b,\hat c}}\in \mathbb{C} {^{1 \times J}}$ is the frequency spectrum of the $h$-th sub control filter. The sub control filter matrix is stated as 
\begin{equation}
\setlength{\abovedisplayskip}{2pt}
\setlength{\belowdisplayskip}{2pt}
{{\bf{c}}^{\hat a,\hat b,\hat c}} = \left[{{\bf{c}}_1^{\hat a,\hat b,\hat c}},{{\bf{c}}_2^{\hat a,\hat b,\hat c}},\ldots,{{\bf{c}}_H^{\hat a,\hat b,\hat c}} \right]^{\rm T}\in \mathbb{R} {^{H \times JL_{\rm c}}},
\end{equation}
where ${{\bf{c}}_h^{\hat a,\hat b,\hat c}}\in \mathbb{R} {^{1 \times JL_{\rm c}}}$ is the time-domain representation of the $h$-th $(h = 1,2,\ldots,H)$ sub control filter and is the basis for generating various control filters.
The generated control filter is computed as the inner product between the sub control filter matrix and the combination weight vector, expressed as
\begin{equation}
\setlength{\abovedisplayskip}{2pt}
\setlength{\belowdisplayskip}{2pt}
{\bf{w}}_{{\rm{gen}}}^{\hat a,\hat b,\hat c} = {\bf{\hat g}} \cdot {\rm{ }}{{\bf{c}}^{\hat a,\hat b,\hat c}} \in \mathbb{R} {^{1 \times JL_{\rm c}}}.
\end{equation}

The soft combination weight vector, with elements ranging from $0$ to $1$, captures the frequency characteristics of the noise source, enabling the generation of a control filter tailored to its spectral content. Notably, the generated control filter is formed by the combination of sub control filters, eliminating the need to train a large number of control filters for different noise types.

\vspace*{-0.3cm}
\subsection{Delayless Noise Control}
\vspace*{-0.1cm}
The pre-trained control filter library development and the CRNN training are performed offline. During online noise control, the proposed SF-GFANC method employs two distinct modules: a co-processor and a real-time controller. The co-processor (e.g., a mobile phone or laptop) executes the CRNN and generates the appropriate control filter at the frame rate, while the real-time controller operates at the sampling rate to perform immediate noise cancellation. This coordinated operation ensures delayless noise control by decoupling real-time processing from the latency introduced by the CRNN. Moreover, this approach minimizes the total electrical processing delay, thereby mitigating the causality bottlenecks that are typically introduced by block-based frequency-domain filtering. The pseudo-code outlining the SF-GFANC noise control procedure is presented in Table~\ref{tabel1}.

\vspace*{-0.3cm}
\subsubsection{Co-processor}
\vspace*{-0.1cm}
In the co-processor, the STFT is applied to each frame of the $J$-channel reference signal. The resulting magnitude and phase maps are concatenated and fed into the CRNN. Based on the predicted distance, elevation, and azimuth indices, the corresponding control filter is retrieved from the pre-trained library, as described in Section \ref{sec:selection}. The selected control filter is decomposed into multiple sub control filters, which are then recombined using the CRNN-predicted combination weights that encode the noise’s frequency characteristics, as described in Section \ref{sec:generation}. The generated control filter is updated to the real-time controller on a frame-by-frame basis. Moreover, the entire control filter generation process is fully data-driven, requiring no prior knowledge or manual intervention.

\begin{table}[!t]
\centering
\small
\caption{Pseudo-code of the SF-GFANC method.}\vspace*{0.1cm}
\begin{tabularx}{\linewidth}{@{}X@{}}
\toprule
\textbf{Initialization:} Control filter vector and combination weights are initialized to zero.\\
\textbf{Input:} a single frame of the $J$-channel reference signal $\mathbf{r}$.\\

\midrule
\textbf{While SF-GFANC on:}\\

\textbf{\# Delayless noise control in the real-time controller (sampling rate):}\\
\hspace*{1em} \textbf{for} each sample of the reference signal \textbf{do}\\
\hspace*{2em} $e(n) = d(n) - s(n)*[{{\bf{w}}_{{\rm{gen}}}^{\hat a,\hat b,\hat c}}(n){\bf{r}}(n)]$ \hfill$\triangleright$ Real-time noise control.\\
\hspace*{1em} \textbf{end for}\\

\textbf{\# Control filter generation in the co-processor (frame rate):}\\
\hspace*{1em} \textbf{for} each frame of the reference signal \textbf{do}\\
\hspace*{2em} $\mathbf{R} \leftarrow \operatorname{Concat}\!\big[\left| {{\rm{STFT}}({\bf{r}})} \right|,\angle {\rm{STFT}}({\bf{r}})\big]$ \hfill$\triangleright$ Channel axis concatenation.\\
\hspace*{2em} $(\hat a,\hat b,\hat c,{\bf{\hat g}}) \leftarrow {\rm{CRNN}}({\bf{R}};{\Theta ^*})$ \hfill$\triangleright$ Spatial-frequency cues estimation.\\
\hspace*{2em} ${{{\bf{w}}^{\hat a,\hat b,\hat c}}}^\prime \leftarrow \hat a, \hat b, \hat c$ \hfill$\triangleright$ Control filter selection.\\
\hspace*{2em} ${{{\bf{c}}^{\hat a,\hat b,\hat c}}}^\prime\leftarrow {{{\bf{w}}^{\hat a,\hat b,\hat c}}}^\prime$ \hfill$\triangleright$ Control filter decomposition.\\
\hspace*{2em} ${{\bf{w}}_{{\rm{gen}}}^{\hat a,\hat b,\hat c}}^\prime = {\bf{\hat g}} \cdot {{\bf{c}}^{\hat a,\hat b,\hat c}}^\prime$ \hfill$\triangleright$ Control filter generation.\\
\hspace*{1em} \textbf{end for}\\

\textbf{\# Control filter update in the real-time controller (frame rate):}\\
\hspace*{1em} \textbf{if} ${\bf{w}}_{{\rm{gen}}}^{\hat a,\hat b,\hat c} \neq {{\bf{w}}_{{\rm{gen}}}^{\hat a,\hat b,\hat c}}^\prime$ \textbf{then}\\
\hspace*{2em} ${\bf{w}}_{{\rm{gen}}}^{\hat a,\hat b,\hat c} \leftarrow {{\bf{w}}_{{\rm{gen}}}^{\hat a,\hat b,\hat c}}^\prime$\\
\hspace*{1em} \textbf{end if}\\

\bottomrule
\end{tabularx}
\label{tabel1}
\vspace*{-0.4cm}
\end{table}

\vspace*{-0.3cm}
\subsubsection{Real-time Controller}
\vspace*{-0.1cm}
In the real-time controller, the generated control filter ${\bf{w}}_{{\rm{gen}}}^{\hat a,\hat b,\hat c}$ is used for real-time noise control at the sampling rate as 
\begin{equation}
\setlength{\abovedisplayskip}{2pt}
\setlength{\belowdisplayskip}{2pt}
e(n) = d(n) - s(n)*[{{\bf{w}}_{{\rm{gen}}}^{\hat a,\hat b,\hat c}}(n){\bf{r}}(n)].
\end{equation}

Unlike traditional adaptive ANC algorithms, the proposed SF-GFANC method eliminates reliance on the error signal for control filter updates. By directly generating the appropriate control filter at the frame level, the system avoids the convergence period typical of adaptive architectures. This enables near-instantaneous adaptation to changing acoustic conditions and reduces the risk of instability. As a result, the system provides a practical solution for suppressing noise sources with diverse 3D locations and frequency characteristics in real-world reverberant environments.

\vspace*{-0.4cm}
\section{Numerical Simulations}
\vspace*{-0.3cm}
\label{simulations}
This section validates the efficacy of the proposed SF-GFANC method through numerical simulations on both simulated and measured acoustic paths across various 3D source locations and noise types. The proposed CRNN is evaluated on unseen noise signals and acoustic environments and compared with a traditional baseline. In addition, SF-GFANC is compared with several representative ANC algorithms to demonstrate its noise reduction performance.

\begin{table}[!t]
\centering
\caption{Parameter settings used in the numerical simulations.}\vspace*{0.1cm}
\small
\begin{tabular}{|c|c|c|}
\hline
\textbf{Variable} & \textbf{Definition} & \textbf{Value} \\
\hline
$J$ & Number of reference microphones & $4$ \\
$c_{\rm{a}}$ & Sound speed in air & $343$ m/s \\
$L_{\rm c}$ & Control filter length & $1024$ \\
$L_{\rm r}$ & Reference signal length & $8000$ \\
-- & Secondary path length & $256$ \\
$F$ & STFT frequency bins & $513$ \\
$T$ & STFT time frames & $64$ \\
$A$ & Number of distance categories & $4$ \\
$B$ & Number of elevation angle categories & $3$ \\
$C$ & Number of azimuth angle categories & $6$ \\
$H$ & Number of sub control filters & $8$ \\
-- & Sampling rate & $16000$ Hz \\
-- & Frame length & $0.5$ s \\
\hline
\end{tabular}
\label{tabel2}
\vspace*{-0.3cm}
\end{table}

\vspace*{-0.3cm}
\subsection{Simulation Setup}
\vspace*{-0.1cm}
The parameter settings for the numerical simulations are summarized in Table \ref{tabel2}. These parameters are determined through validation-based tuning to balance performance and computational efficiency, and can be flexibly adjusted according to specific system requirements. Specifically, a tetrahedral microphone setup is used as the reference microphone array, which consists of four cardioid microphones arranged with a diameter of $0.025$ $\text{m}$ and follows the configuration of the \textit{Sennheiser AMBEO VR Mic}. This setup enables effective capture of the 3D spatial information of the noise source in a compact form and preserves the necessary spatial cues required for the proposed task \cite{kushwaha2023sound}. Notably, given the small aperture of the reference microphone array and the source distances considered, the far-field assumption holds in the simulation setting.

\vspace*{-0.3cm}
\subsubsection{Pre-trained Control Filter Library}
\begin{figure}[!t]
\centering
\includegraphics[width=3.9in]{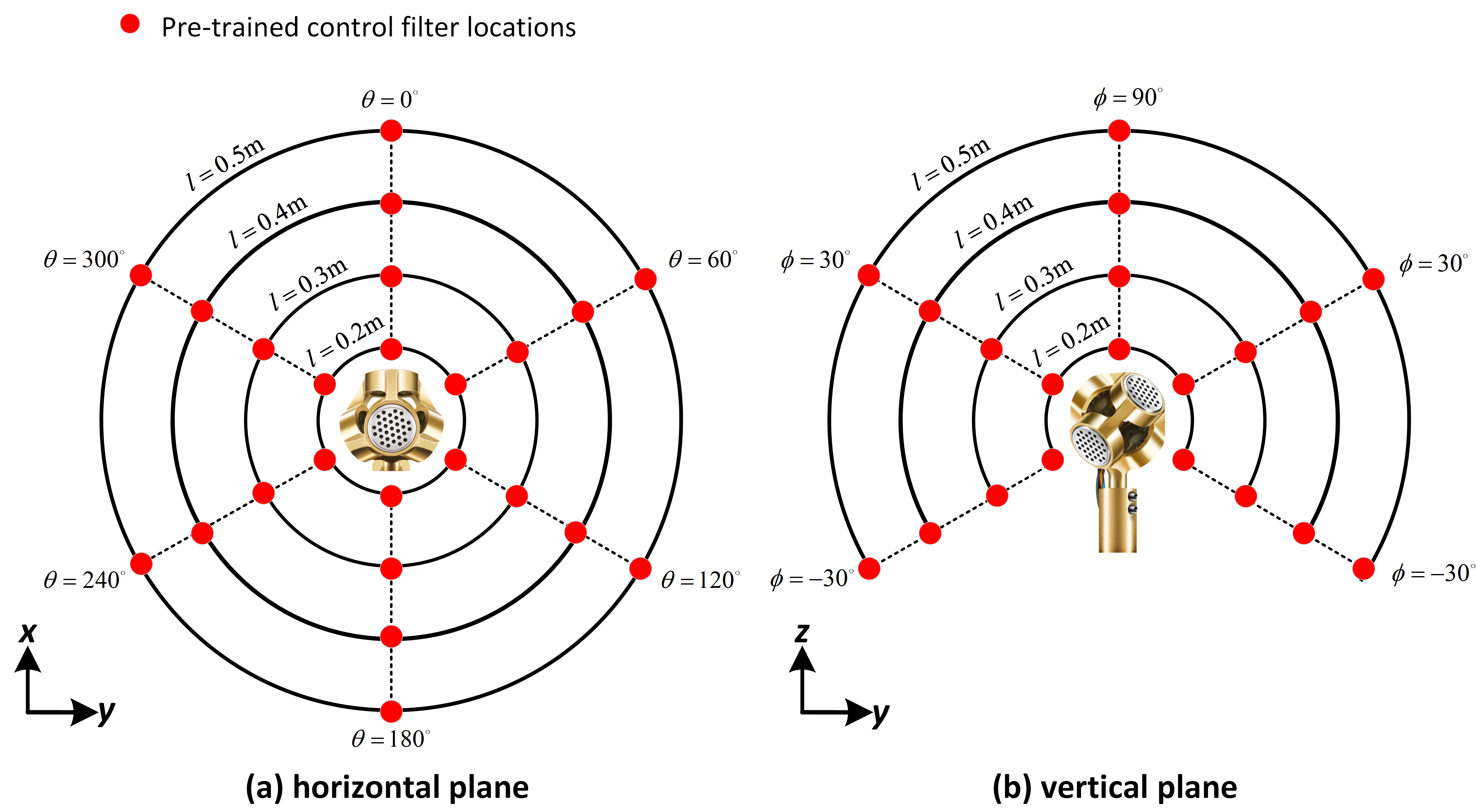}\vspace*{-0.4cm}
\caption{Illustration of the pre-trained control filter locations: (a) the horizontal plane view and (b) the vertical plane view.}
\label{fig_4}
\end{figure}

\vspace*{-0.1cm}
As illustrated in Fig.~\ref{fig_4}, the spatial region surrounding the reference microphone array is uniformly discretized into multiple grid points to construct the pre-trained control filter library. Specifically, four distance classes $l \in [0.2,0.3,0.4,0.5]$ $\text{m}$ are measured from the center of the reference microphone array. Three elevation angle classes $\phi  \in [- 30,30,90]^\circ$ are used, where $90^\circ$ corresponds to the positive $z$-axis direction and six azimuth angle classes $\theta  \in [0,60,120,180,240,300]^\circ$ are defined relative to the positive $x$-axis. At each spatial grid point, a control filter is pre-trained using the FxLMS algorithm with broadband noise in the $20-2020$ Hz range, covering the low-frequency band typically targeted by ANC systems \cite{wang2025transferable}. In total, $52$ control filters are pre-trained and stored in the library.

\begin{figure}[!t]
\centering
\includegraphics[width=3.9in]{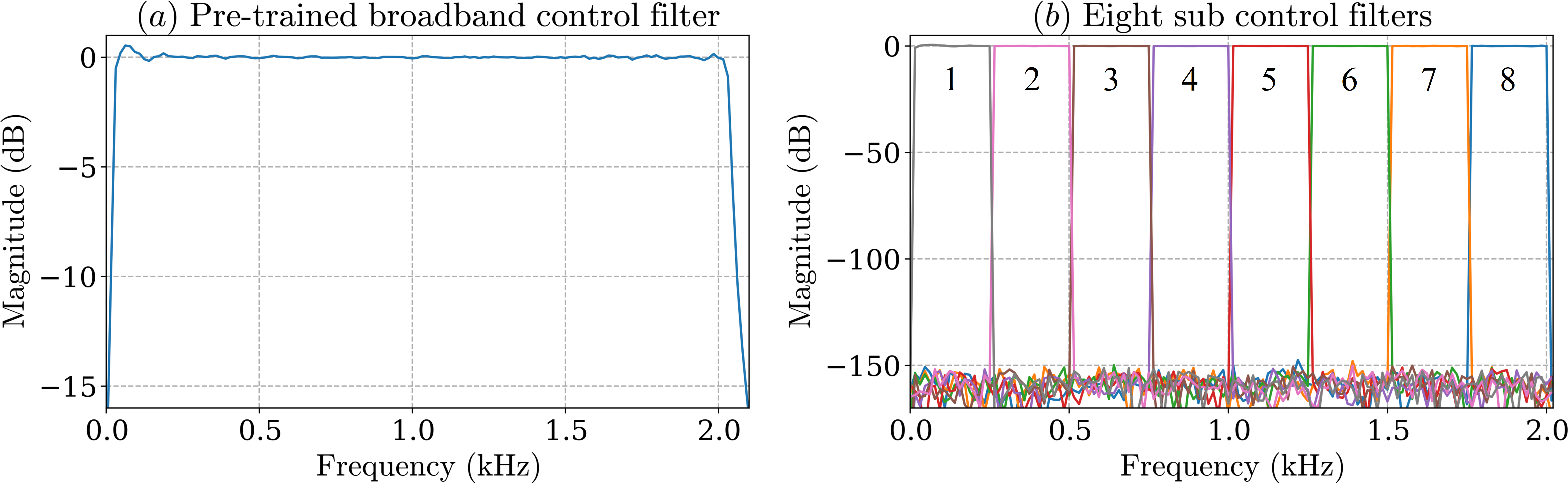}\vspace*{-0.4cm}
\caption{Magnitude response of (a) pre-trained broadband control filter and (b) eight sub control filters.}
\label{fig_5}
\vspace*{-0.3cm}
\end{figure}

\vspace*{-0.3cm}
\subsubsection{Sub Control Filters}
\vspace*{-0.1cm}
The pre-trained broadband control filter is decomposed into eight sub control filters as described in Section~\ref{sec:generation}. Under synthetic acoustic paths, the frequency spectra of both the pre-trained broadband control filter and its eight sub control filters are illustrated in Fig.~\ref{fig_5}. The broadband spectrum is evenly divided, with each sub control filter covering a distinct and equal portion of the overall frequency range. This decomposition enables the generation of a control filter whose frequency content closely matches that of the noise.

\vspace*{-0.3cm}
\subsubsection{CRNN Configurations}
\vspace*{-0.1cm}
The CRNN is trained using the Adam optimizer, with the initial learning rate set to ${10^{ - 3}}$. To optimize training, the learning rate is adaptively adjusted based on the loss of the validation dataset. Specifically, if the validation loss does not improve for five consecutive epochs, the learning rate is reduced by a factor of $0.5$. A minimum learning rate of ${10^{ - 6}}$ is enforced to prevent it from decreasing excessively. Early stopping is applied with a patience of $15$ epochs to avoid overfitting while ensuring sufficient training.

\vspace*{-0.3cm}
\subsection{Datasets Generation}
\vspace*{-0.1cm}
\label{sectiondata}
The training, validation, and testing datasets for the CRNN are generated by simulating the multichannel reference signals through convolving noise signals with RIRs. The noise signals include both synthesized and real-world noises: the synthesized noises are generated as bandlimited white noise with random bandwidths (minimum $60$ Hz) within the $20-2020$ Hz range, while the real-world noises are sourced from the \textit{UrbanSound8K} dataset \cite{salamon2014dataset}. The RIRs are generated using the \textit{gpuRIR} library \cite{diaz2021gpurir}, which implements the image source method \cite{allen1979image}. Specifically, a variety of RIRs corresponding to different 3D noise source locations are used to simulate spatial diversity. For each instance, the source-to-array distance is randomly selected from the range $[0.1, 0.6]$ $\text{m}$, the elevation angle from $[-60, 90]^\circ$ and the azimuth angle from $[0, 360]^\circ$. The array-to-surface distance is maintained at more than one meter to avoid excessive wall reflections. The ground-truth labels for distance, elevation angle, and azimuth angle are assigned by mapping each simulated location to its nearest class as defined in Fig.~\ref{fig_4}. The corresponding optimal soft combination weights are computed following the method introduced in \cite{luo2023delayless}.

\begin{table}[!t]
    \centering
    \small
    \caption{Configurations for training, validation and testing datasets.}\vspace*{0.1cm}
    \begin{tabular}{cc}
        \toprule
        \multicolumn{2}{c}{\textbf{Training and Validation Datasets}} \\
        \midrule
        Noise signal & Synthesized noises \& real noises \\
        Room size ($\text{m}^3$) & $\rm{R}_1$: $(6\times4\times3)$, $\rm{R}_2$: $(12\times8\times3.5)$, $\rm{R}_3$: $(16\times14\times4)$ \\
        Array positions & $8$ arbitrary positions in each room \\
        $\rm{RT}_{60}$ (s) & $\rm{R}_1$: $0.1$, $0.2$, $0.3$; $\rm{R}_2$: $0.4, 0.5, 0.6$; $\rm{R}_3$: $0.7, 0.8, 0.9$ \\
        SNR (dB) & Uniformly sampled from $10$ to $50$ \\
        \midrule
        \multicolumn{2}{c}{\textbf{Testing Dataset}} \\
        \midrule
        Noise signal & Synthesized noises \& real noises \\
        Room size ($\text{m}^3$) & ${\rm{R}_1}^\prime$: $(7\times5\times3)$; ${\rm{R}_2}^\prime$: $(11\times9\times3.2)$; ${\rm{R}_3}^\prime$: $(15\times13\times4.2)$ \\
        Array positions & $4$ arbitrary positions in each room \\
        $\rm{RT}_{60}$ (s) & ${\rm{R}_1}^\prime$: $0.17$, ${\rm{R}_2}^\prime$: $0.48$, ${\rm{R}_3}^\prime$: $0.83$ \\
        SNR (dB) & $10$, $20$, $30$, $40$, $50$ \\
        \bottomrule
    \end{tabular}
    \label{table3}
    \vspace*{-0.3cm}
\end{table}
 
To enhance the CRNN's robustness across diverse acoustic environments, the training dataset incorporates variations of the RIRs generated under different room sizes, array positions and reverberation time ($\rm{RT}_{60}$). Specifically, longer $\rm{RT}_{60}$ values are paired with larger room volumes to reflect more natural reverberation distributions \cite{garcia2022binaural}. Additionally, a range of signal-to-noise ratio (SNR) levels is applied to ensure robust model performance in noisy conditions. A summary of the configurations of the datasets is provided in Table \ref{table3}. Importantly, the testing dataset includes both noise signals and acoustic environments that are entirely distinct from those used during training, ensuring a fair evaluation of the model’s generalization capability. We simulate three additional rooms and sample four array positions per room, using unseen $\mathrm{RT}_{60}$ values. In addition, five different levels of white noise are included, yielding $15$ testing subsets, one for each room–SNR combination, thereby enabling a controlled performance comparison.

In total, the datasets include $46080$ training samples ($38400$ synthesized and $7680$ real), $5760$ validation samples ($4800$ synthesized and $960$ real), and $4800$ test samples ($4000$ synthesized and $800$ real) per room–SNR subset. Each sample is a four-channel reference signal with a duration of $0.5$ seconds.

\vspace*{-0.3cm}
\subsection{Performance on Simulated Acoustic Paths}
\begin{figure}[!t]
\centering
\includegraphics[width=3.9in]{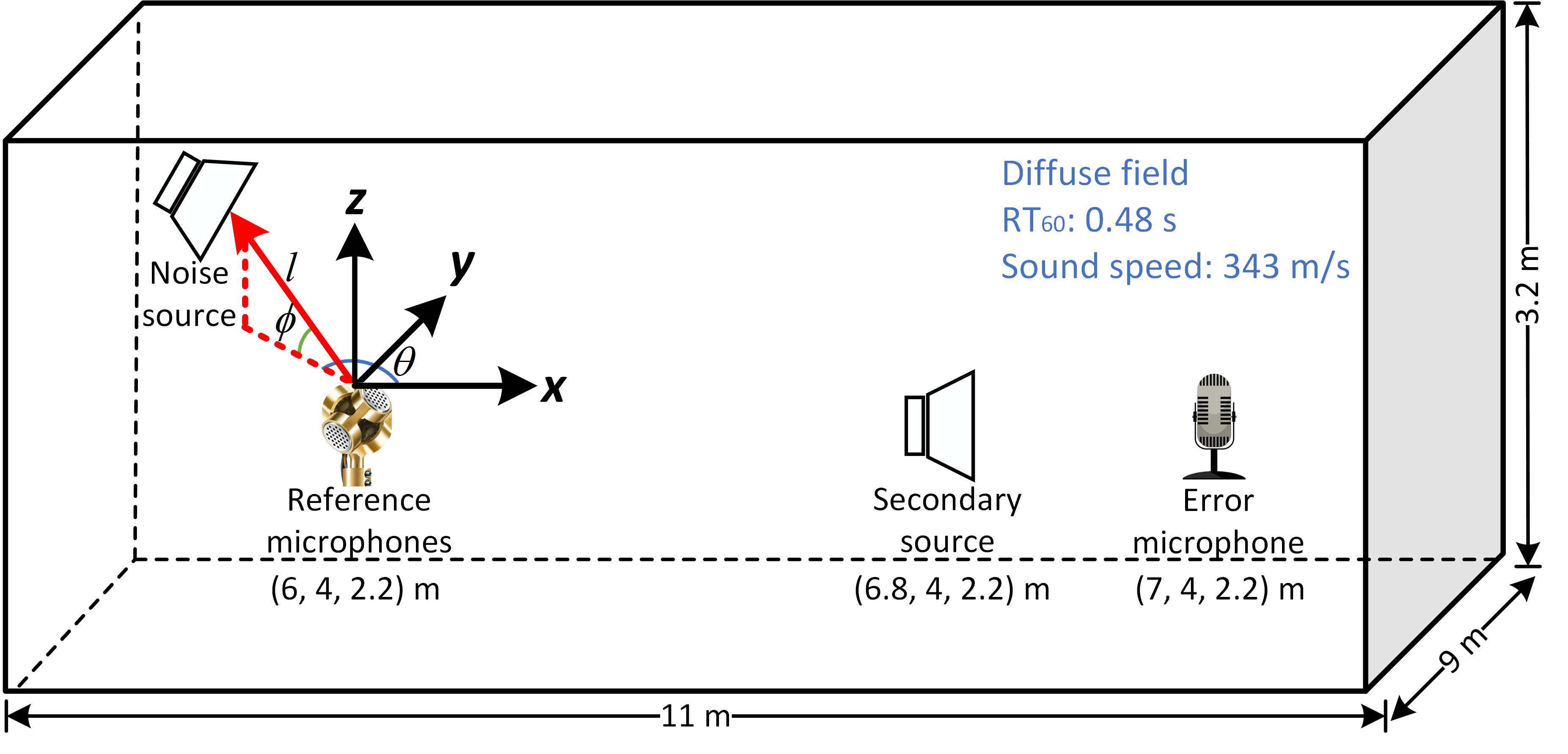}\vspace*{-0.4cm}
\caption{Simulation arrangement of the multi-reference ANC system with four reference microphones, one secondary source, and one error microphone.}
\label{fig_6}
\vspace*{-0.3cm}
\end{figure}

\vspace*{-0.1cm}
\label{Simulated Acoustic Paths}
This section evaluates the performance of the proposed SF-GFANC method using simulated RIRs \cite{diaz2021gpurir}. First, the CRNN’s effectiveness is assessed in terms of distance, elevation angle, and azimuth angle classifications, as well as combination weights regression, with a traditional DoA estimation method used as a baseline for comparison. In addition, the complexity of the CRNN is analyzed to assess its efficiency for practical deployment. Second, the SF-GFANC method is applied to attenuate both broadband and real-world noise sources at different 3D locations to evaluate its noise reduction performance. The noise reduction simulations are conducted in an unseen environment using a setup comprising four reference microphones, one secondary source, and one error microphone, as illustrated in Fig.~\ref{fig_6}, with the SNR level fixed at $50$ dB for all simulations. The proposed SF-GFANC method is compared with several representative ANC methods, including the FxLMS algorithm, the SFANC method \cite{shi2022selective}, the GFANC method \cite{luo2023deep}, and the frequency-direction aware SFANC (FD-SFANC) method \cite{luo2025doa}. It is worth noting that the CRNN is used directly for estimating spatial and frequency cues without retraining, while only the pre-trained control filter library needs to be updated based on the system-specific acoustic paths.

\vspace*{-0.3cm}
\subsubsection{Effectiveness of the CRNN}
\vspace*{-0.1cm}
\label{CRNN effectiveness}
For each room and SNR level, Table~\ref{table4} reports the classification accuracies for distance, elevation, and azimuth, together with the MSE of the combination weights. For each SNR level, results are averaged over the four array positions considered in each room. The widely used SRP using phase transform (SRP-PHAT) method \cite{dibiase2000high} is adopted as a baseline for elevation and azimuth estimation. It can be seen from the results that the proposed CRNN consistently achieves high 3D localization accuracy while maintaining a low MSE loss. As the SNR level decreases from $50$ dB to $10$ dB, the performance degradation is marginal. These results demonstrate the generalization capability of the CRNN when applied to unseen noise types and reverberant environments. Conversely, the SRP-PHAT exhibits lower elevation and azimuth accuracies and is more sensitive to SNR variations. This gap can be attributed to room reverberation, the small aperture of the microphone array, and the low-frequency nature of the noise signals. Furthermore, unlike traditional DoA methods that fail to provide source distance or frequency cues, the multi-task CRNN estimates the noise’s 3D location and spectral content within a unified framework, enabling control filter generation that exploits both 3D spatial and frequency information.

Regarding model complexity, the proposed CRNN offers high efficiency, containing only $0.24$ million parameters and a computational load of $121.26$ million multiply–accumulate operations (MACs), which makes it suitable for deployment on resource-constrained co-processors. Importantly, this computation is entirely offloaded to the co-processor and executed at the frame rate, thereby decoupling it from the real-time controller operating at the sampling rate.

\begin{table}[!t]
\centering
\scriptsize
\caption{Performance of the proposed CRNN under three unseen environments at different SNR levels, with SRP-PHAT as a baseline for elevation and azimuth estimation.}\vspace*{0.1cm}
\begin{tabular}{cccccccc}
\toprule
\textbf{Metric} & \textbf{Env.} & \textbf{Method} & \multicolumn{5}{c}{\textbf{SNR (dB)}}\\
\cmidrule(lr){4-8}
 &  &  & \textbf{10} & \textbf{20} & \textbf{30} & \textbf{40} & \textbf{50}\\
\midrule

\multirow{3}{*}{Distance accuracy (\%)}
 & ${\rm R_1}'$ & CRNN & 94.27 & 94.69 & 95.21 & 95.02 & 94.46\\
 & ${\rm R_2}'$ & CRNN & 91.83 & 92.10 & 93.04 & 92.96 & 92.69\\
 & ${\rm R_3}'$ & CRNN & 92.87 & 94.33 & 94.17 & 94.40 & 93.96\\
\midrule

\multirow{6}{*}{Elevation accuracy (\%)}
 & \multirow{2}{*}{${\rm R_1}'$} & SRP-PHAT       & 79.60 & 83.27 & 83.17 & 85.56 & 87.52 \\
 &                              & \textbf{CRNN}     & \textbf{92.44} & \textbf{94.52} & \textbf{94.46} & \textbf{95.35} & \textbf{95.02}\\
 & \multirow{2}{*}{${\rm R_2}'$} & SRP-PHAT       & 77.65 & 81.31 & 80.40 & 82.60 & 84.27 \\
 &                              & \textbf{CRNN}     & \textbf{90.69} & \textbf{93.42} & \textbf{93.17} & \textbf{94.02} & \textbf{93.90}\\
 & \multirow{2}{*}{${\rm R_3}'$} & SRP-PHAT       & 78.60 & 81.85 & 81.94 & 84.73 & 85.15 \\
 &                              & \textbf{CRNN}     & \textbf{91.40} & \textbf{94.15} & \textbf{94.04} & \textbf{94.58} & \textbf{94.81}\\
\midrule

\multirow{6}{*}{Azimuth accuracy (\%)}
 & \multirow{2}{*}{${\rm R_1}'$} & SRP-PHAT       & 65.01 & 72.65 & 75.91 & 78.32 & 79.66 \\
 &                              & \textbf{CRNN}     & \textbf{97.23} & \textbf{98.10} & \textbf{97.67} & \textbf{97.71} & \textbf{97.83}\\
 & \multirow{2}{*}{${\rm R_2}'$} & SRP-PHAT       & 60.29 & 69.05 & 69.90 & 71.44 & 75.22 \\
 &                              & \textbf{CRNN}     & \textbf{95.85} & \textbf{96.15} & \textbf{96.25} & \textbf{96.19} & \textbf{96.19}\\
 & \multirow{2}{*}{${\rm R_3}'$} & SRP-PHAT       & 61.83 & 70.61 & 72.99 & 76.02 & 77.23 \\
 &                              & \textbf{CRNN}     & \textbf{97.10} & \textbf{97.33} & \textbf{97.81} & \textbf{97.71} & \textbf{97.04} \\
\midrule

\multirow{3}{*}{Weights MSE}
 & ${\rm R_1}'$ & CRNN & 0.0531 & 0.0406 & 0.0594 & 0.0496 & 0.0445\\
 & ${\rm R_2}'$ & CRNN & 0.0523 & 0.0386 & 0.0525 & 0.0472 & 0.0459\\
 & ${\rm R_3}'$ & CRNN & 0.0527 & 0.0397 & 0.0551 & 0.0465 & 0.0447\\

\bottomrule
\end{tabular}
\label{table4}
\vspace*{-0.35cm}
\end{table}

\vspace*{-0.3cm}
\subsubsection{Importance of 3D Localization}
\vspace*{-0.1cm}
To investigate the noise reduction performance of the SF-GFANC method under different 3D noise source locations, a $100-700$ Hz broadband noise is positioned at $(l = 0.3 \ \text{m},\phi  = 30^\circ, \theta  = 0^\circ )$, $(l = 0.2 \ \text{m},\phi  = -30^\circ, \theta  = 0^\circ )$ and $(l = 0.2 \ \text{m},\phi  = 30^\circ, \theta  = 120^\circ )$ relative to the reference microphone array. The PSDs and averaged noise reduction (NR) levels of the residual noise after applying the FxLMS, SFANC, GFANC, and SF-GFANC methods are shown in Fig. \ref{fig_7}. PSDs are computed over the entire duration of the signal, while average NR levels are calculated every $0.5$ seconds to align with the control filter update rate. The averaged NR level, measured in dB, is defined as 
\begin{equation}
\setlength{\abovedisplayskip}{2pt}
\setlength{\belowdisplayskip}{2pt}
{\rm{NR}} = 10{\log _{10}}\frac{{\sum\nolimits_{n = 1}^N {{d^2}(n)} }}{{\sum\nolimits_{n = 1}^N {{e^2}(n)} }},
\end{equation}
where $N$ denotes the length of the signal. The SFANC method employs $15$ pre-trained control filters, each corresponding to a specific frequency range: $20-2020$ Hz, $20-1020$ Hz, $1020-2020$ Hz, $20-520$ Hz, $520-1020$ Hz, $1020-1520$ Hz, $1520-2020$ Hz, $20-270$ Hz, $270-520$ Hz, $520-770$ Hz, $770-1020$ Hz, $1020-1270$ Hz, $1270-1520$ Hz, $1520-1770$ Hz, and $1770-2020$ Hz. In contrast, the GFANC method utilizes eight sub control filters that uniformly partition the $20-2020$ Hz frequency range, as illustrated in Fig.~\ref{fig_5}. For both the SFANC and GFANC methods, the noise source location is assumed to be fixed at $(l = 0.2 \ \text{m},\phi  = 30^\circ, \theta  = 0^\circ )$. The step size of the FxLMS algorithm is set to $1 \times {10^{ - 4}}$. The corresponding CRNN outputs are summarized in Table~\ref{table:CRNNoutputs}, indicating that the SF-GFANC method generates appropriate control filters for noise sources at different 3D locations.

\begin{figure}[!t]
\centering
\includegraphics[width=3.9in]{Figure7.png}\vspace*{-0.25cm}
\caption{
Noise reduction performance achieved by different ANC algorithms for a $100$–$700$ Hz broadband noise located at different 3D locations.}
\label{fig_7}
\vspace*{-0.3cm}
\end{figure}

\begin{table}[!t]
\centering
\caption{CRNN outputs for various noise types and 3D source locations.}\vspace*{0.1cm}
\setlength{\tabcolsep}{3pt}
\renewcommand{\arraystretch}{1.3} 
\scriptsize

\begin{tabular}{lccc|cccc}
\hline 
\multicolumn{4}{c|}{\textbf{Ground Truth}} & \multicolumn{4}{c}{\textbf{CRNN Outputs}} \\
\hline 
\textbf{Noise Type} & \textbf{Dist.} & \textbf{Elev.} & \textbf{Azim.} &
\textbf{Dist.} & \textbf{Elev.} & \textbf{Azim.} & \textbf{Combination Weights $(\bf{\hat g})$} \\
\hline 

$100$--$700$ Hz & 0.3 \text{m} & $30^\circ$  & $0^\circ$   & 0.3 \text{m}& $30^\circ$  & $0^\circ$   & $[1.00,1.00,1.00,0.13,0.04,0.02,0.02,0.02]$\\
$100$--$700$ Hz & 0.2 \text{m}& $-30^\circ$ & $0^\circ$   & 0.2 \text{m}& $-30^\circ$ & $0^\circ$   & $[1.00,1.00,1.00,0.10,0.03,0.02,0.02,0.01]$\\
$100$--$700$ Hz & 0.2 \text{m}& $30^\circ$  & $120^\circ$ & 0.2 \text{m}& $30^\circ$  & $120^\circ$ & $[1.00,1.00,1.00,0.12,0.04,0.03,0.03,0.02]$\\
$50$--$450$ Hz  & 0.3 \text{m}& $30^\circ$  & $60^\circ$  & 0.3 \text{m}& $30^\circ$  & $60^\circ$  & $[1.00,1.00,0.11,0.05,0.03,0.03,0.04,0.03]$\\
$50$--$450$ Hz  & 0.2 \text{m}& $-30^\circ$ & $120^\circ$ & 0.2 \text{m}& $-30^\circ$ & $120^\circ$ & $[1.00,1.00,0.13,0.06,0.03,0.03,0.03,0.02]$\\
Vacuum           & 0.22 \text{m}& $20^\circ$  & $10^\circ$  & 0.2 \text{m}& $30^\circ$  & $0^\circ$   & $[0.96,1.00,0.99,0.91,0.12,0.02,0.02,0.02]$\\
Washing machine  & 0.22 \text{m}& $20^\circ$  & $10^\circ$  & 0.2 \text{m}& $30^\circ$  & $0^\circ$   & $[1.00,0.91,0.61,0.22,0.05,0.03,0.02,0.02]$\\
Washing machine  & 0.38 \text{m}& $20^\circ$  & $10^\circ$  & 0.4 \text{m}& $30^\circ$  & $0^\circ$   & $[1.00,0.92,0.65,0.32,0.05,0.03,0.02,0.01]$\\
Washing machine  & 0.38 \text{m}& $-20^\circ$ & $10^\circ$  & 0.4 \text{m}& $-30^\circ$ & $0^\circ$   & $[1.00,0.93,0.72,0.32,0.05,0.02,0.02,0.02]$\\
Washing machine  & 0.38 \text{m}& $-20^\circ$ & $70^\circ$  & 0.4 \text{m}& $-30^\circ$ & $60^\circ$  & $[1.00,0.96,0.75,0.44,0.08,0.05,0.04,0.02]$\\
Compressor       & 0.47 \text{m}& $40^\circ$  & $310^\circ$ & 0.5 \text{m}& $30^\circ$  & $300^\circ$ & $[0.98,1.00,0.98,0.39,0.02,0.02,0.01,0.01]$\\
Hand drill       & 0.54 \text{m}& $-20^\circ$ & $250^\circ$ & 0.5 \text{m}& $-30^\circ$ & $240^\circ$ & $[0.99,0.90,0.71,0.77,0.41,0.64,0.24,0.24]$\\

\hline 
\end{tabular}
\label{table:CRNNoutputs}
\vspace*{-0.3cm}
\end{table}

As shown in Fig. \ref{fig_7}, the proposed SF-GFANC method achieves an averaged NR level over $10$ dB after the first $0.5$ seconds, outperforming the FxLMS algorithm in terms of both response time and overall noise reduction performance. This gap mainly arises because the FxLMS controller has not yet converged to a steady state within this interval, whereas SF-GFANC can generate and apply an appropriate control filter at the frame level. However, the control filters used in the SFANC and GFANC methods are determined exclusively based on the frequency characteristics of the noise source. Consequently, when the noise source deviates in any spatial parameter from the pre-trained location, the noise reduction performance of the SFANC and GFANC methods degrades significantly and may even result in noise amplification. In comparison, the SF-GFANC method generates control filters that are better matched to the noise source by jointly incorporating both 3D spatial and frequency information, thereby maintaining effective noise reduction performance across diverse source locations. These findings are consistent with the theoretical analysis presented in Section \ref{theory}, highlighting the importance of jointly estimating the distance, elevation angle, and azimuth angle of the noise source in reverberant environments.

\begin{figure}[!t]
\centering
\includegraphics[width=3.9in]{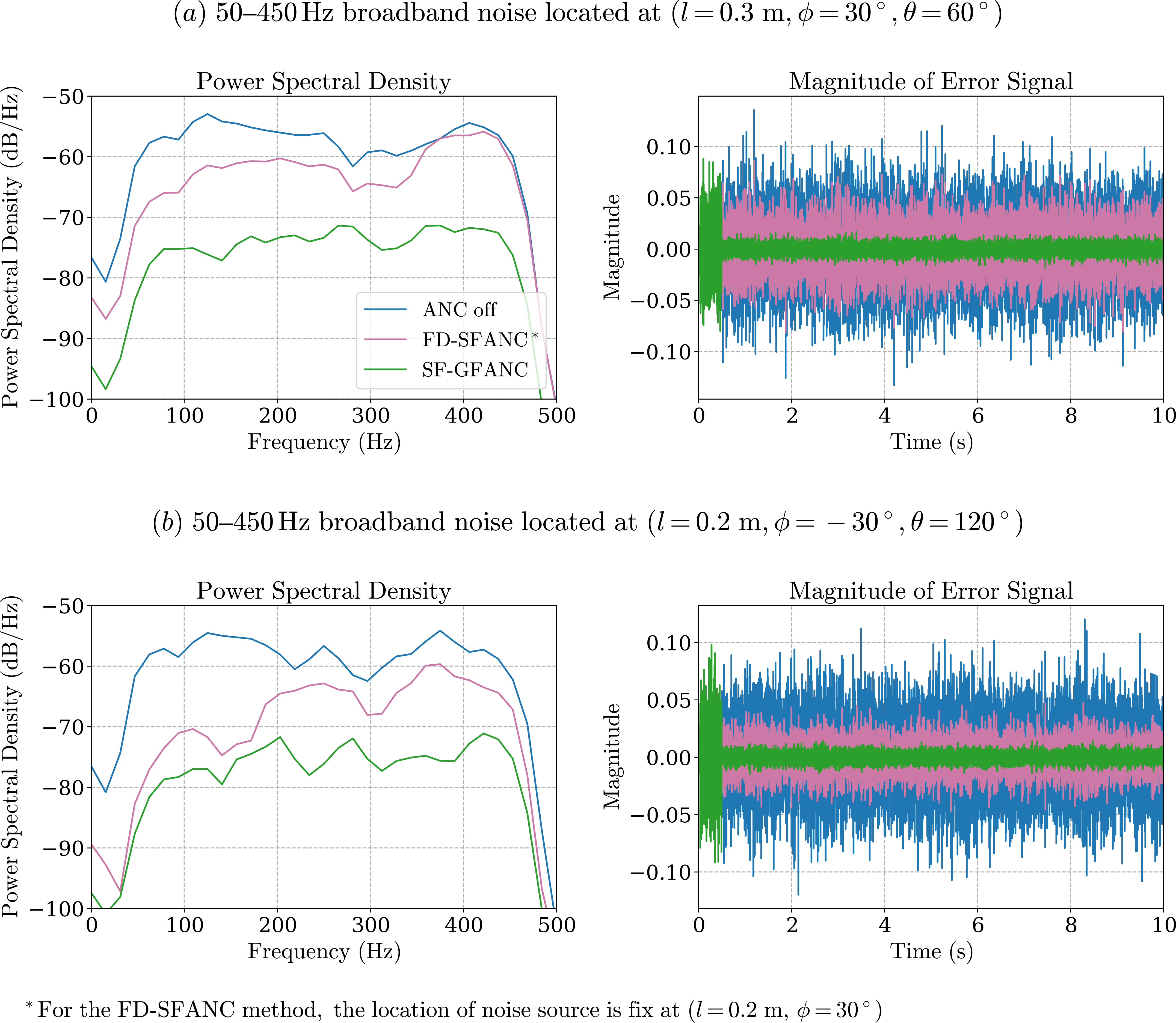}\vspace*{-0.25cm}
\caption{Comparison of noise reduction performance between the FD-SFANC \cite{luo2025doa} and the proposed SF-GFANC for a $50-450$ Hz broadband noise located at different 3D locations.}
\label{fig_8}
\vspace*{-0.3cm}
\end{figure}

\vspace*{-0.3cm}
\subsubsection{Comparison with the FD-SFANC Method}
\vspace*{-0.1cm}
In this subsection, the proposed SF-GFANC method is compared with the FD-SFANC method \cite{luo2025doa}, which selects the most suitable control filter based on the frequency and azimuth angle characteristics of the noise source. For the FD-SFANC method, the noise source is assumed to remain fixed in the plane $(l = 0.2 \ \text{m}, \phi = 30^\circ)$ relative to the reference microphone array. Fig.~\ref{fig_8} compares the noise reduction performance of the FD-SFANC and SF-GFANC methods on a $50-450$ Hz broadband noise located at $(l = 0.3 \ \text{m},\phi  = 30^\circ, \theta  = 60^\circ )$ and $(l = 0.2 \ \text{m},\phi  = -30^\circ, \theta  = 0^\circ )$. As shown in Table~\ref{table:CRNNoutputs}, the CRNN accurately predicts both the noise type and its 3D location. The results in Fig.~\ref{fig_8} demonstrate that the SF-GFANC method achieves superior performance, providing approximately $15$ dB of noise reduction for sources at different 3D locations. In contrast, the FD-SFANC method exhibits limited noise reduction, as it relies solely on the azimuth angle cues for control filter selection while neglecting the distance and elevation angle cues, both of which are critical for effective noise control in reverberant environments.

\begin{figure}[!t]
\centering
\includegraphics[width=4.35in]{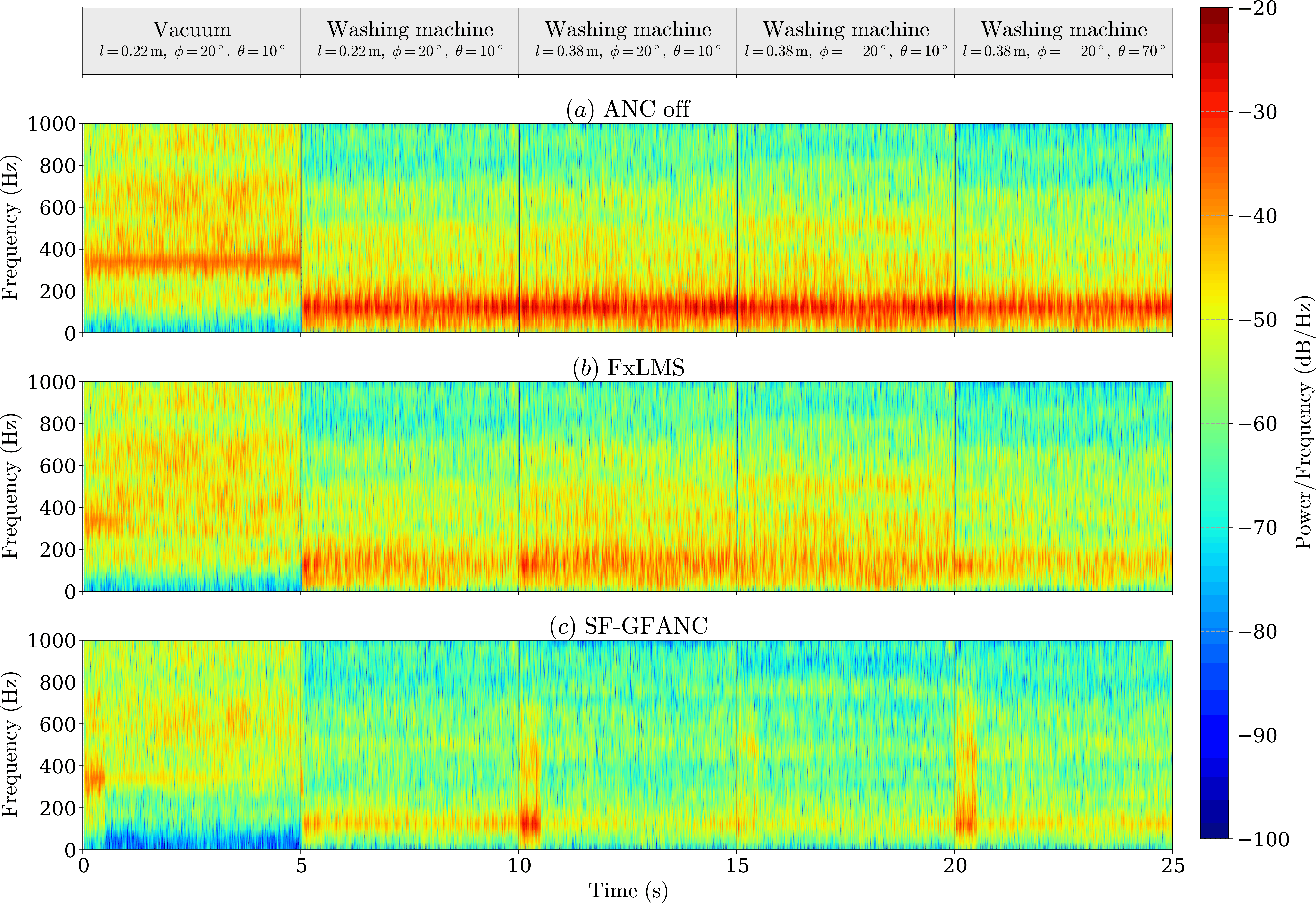}\vspace*{-0.4cm}
\caption{Spectrograms of the error signal for real-world noises with time-varying types and 3D locations: (a) ANC off, (b) FxLMS algorithm, and (c) proposed SF-GFANC method.}
\label{fig_9}
\vspace*{-0.3cm}
\end{figure}

\vspace*{-0.3cm}
\subsubsection{Time-Varying Real-World Noise Cancellation}
\vspace*{-0.1cm}
To evaluate the noise reduction performance of the proposed SF-GFANC method on real-world noises with time-varying 3D location and frequency content, vacuum and washing machine noise segments, each at different 3D locations and lasting five seconds, are concatenated to form a $25$-second test signal. As shown in Table~\ref{table:CRNNoutputs}, the CRNN continuously predicts the appropriate spatial classes, even for source locations not included in the pre-trained control filter library. Fig.~\ref{fig_9} shows the spectrograms of the residual error signals attenuated by the FxLMS algorithm and the SF-GFANC method. It can be observed that the SF-GFANC rapidly adapts to variations in both source location and frequency content, requiring only $0.5$ seconds to update the control filter accordingly. In contrast, the FxLMS algorithm responds more slowly to such variations due to its dependence on gradual convergence. Moreover, the SF-GFANC method consistently achieves higher noise reduction performance than the FxLMS algorithm, particularly in the initial moments following transitions in noise type or location. These findings suggest that slight deviations of the noise source from the pre-trained control filter location lead to minor changes in the control filter, allowing the system to maintain a certain level of noise reduction.

\begin{figure}[!t]
\centering
\includegraphics[width=3.5in]{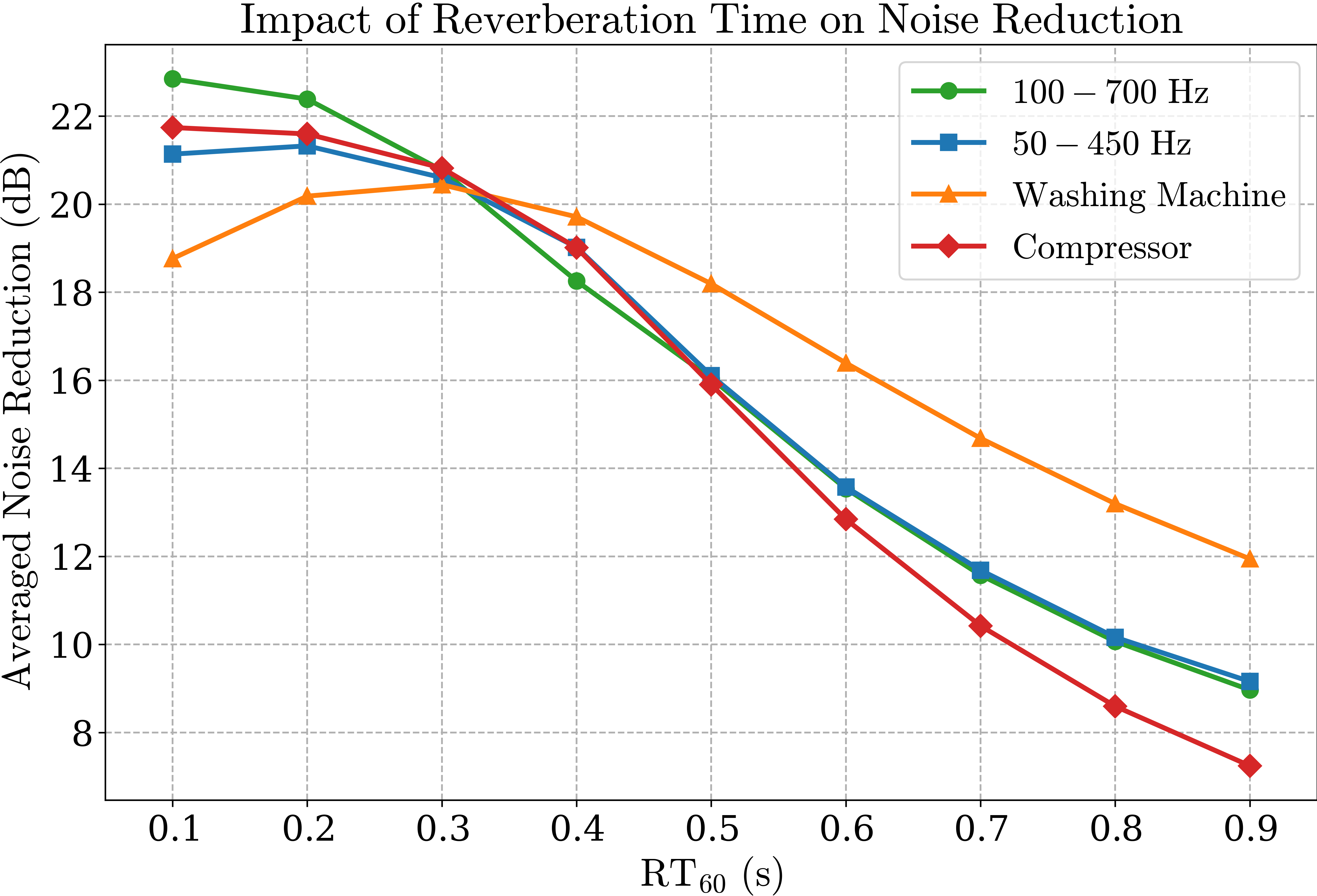}\vspace*{-0.4cm}
\caption{Averaged noise reduction versus reverberation time across various noise types.}
\label{fig_10}
\vspace*{-0.4cm}
\end{figure}

\vspace*{-0.3cm}
\subsubsection{Impact of Reverberation Time}
\vspace*{-0.1cm}
To systematically investigate the relationship between noise reduction performance and reverberation time, a controlled parametric evaluation is conducted. The $\text{RT}_{60}$ is varied from $0.1$ s to $0.9$ s in increments of $0.1$ s, with four types of noise sources positioned at a fixed location ($l=0.2$ m, $\phi=30^\circ$, $\theta=120^\circ$). Fig.~\ref{fig_10} illustrates the averaged NR curves versus $\text{RT}_{60}$ for the proposed SF-GFANC method across the four noise sources. As shown, the method exhibits a clear and anticipated downward trend in noise reduction as $\text{RT}_{60}$ increases. This occurs because strong reverberation induces complex multi-path propagation, which fundamentally reduces the coherence between the reference signal and the disturbance. This analysis confirms that severe reverberation poses an inherent bottleneck for ANC performance. Nevertheless, because the proposed CRNN is trained on highly diverse reverberant datasets (covering $\text{RT}_{60}$ ranges from $0.1$ s to $0.9$ s), it remains robust in accurately predicting spatial and frequency cues despite the highly diffuse sound field. Consequently, the SF-GFANC method continues to generate the appropriate control filters, maintaining an average NR of over $7$ dB even at a high $\text{RT}_{60}$ of $0.9$ s.

\begin{figure}[!t]
\centering
\includegraphics[width=3in]{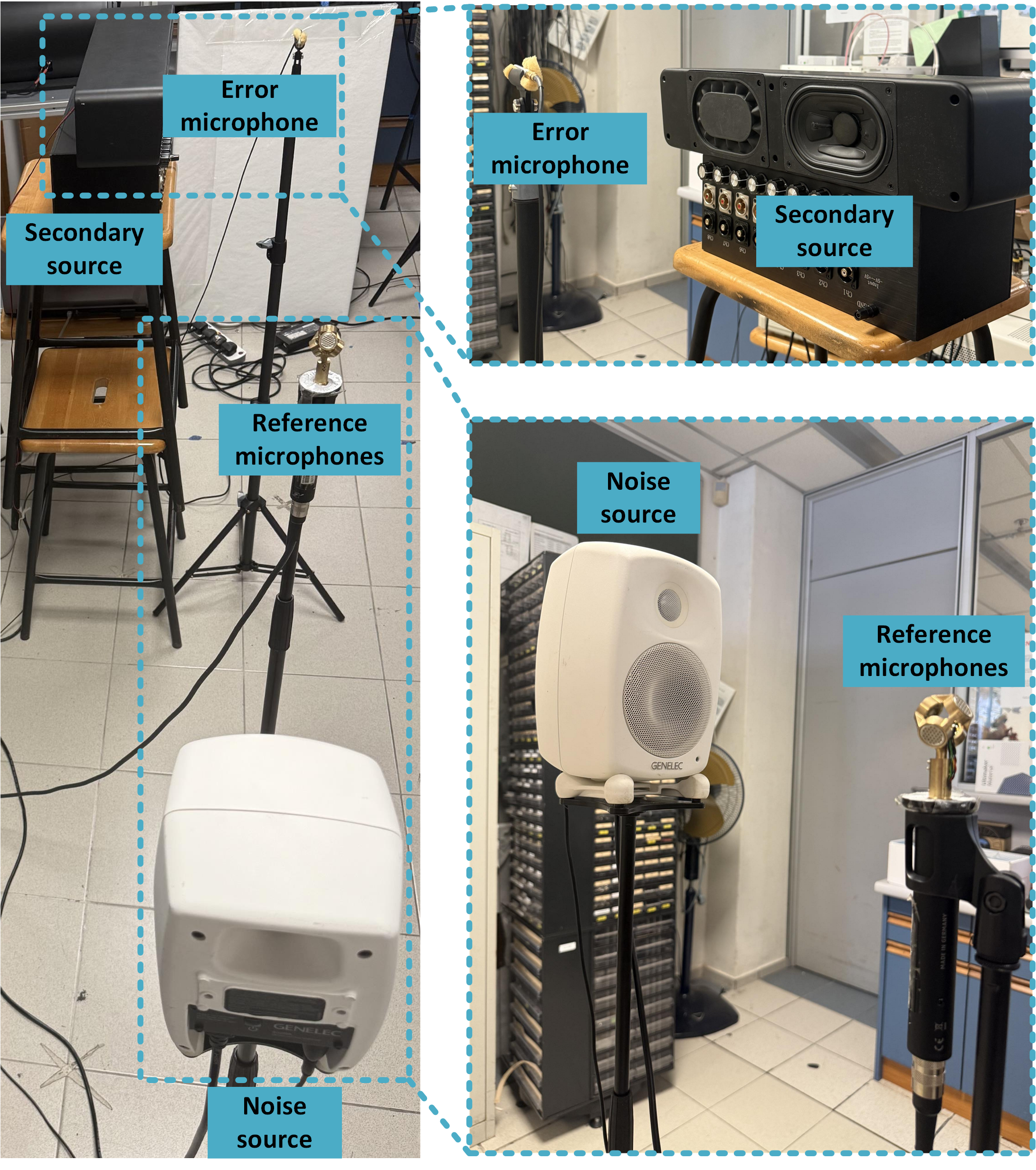}\vspace*{-0.4cm}
\caption{Placement of the reference microphones, secondary source and error microphone for acoustic path measurement.}
\label{fig_11}
\vspace*{-0.3cm}
\end{figure}

\vspace*{-0.3cm}
\subsection{Performance on Measured Acoustic Paths}
\vspace*{-0.1cm}
This section evaluates the performance of the proposed SF-GFANC method using measured acoustic paths. The measurements were conducted in a room with dimensions $8 \times 5.2 \times 2.6$ $\text{m}^3$ and $\rm{RT}_{60}=0.27$ s. A multi-reference ANC system consisting of four reference microphones, one secondary source, and one error microphone was configured at the same height, with the setup illustrated in Fig.~\ref{fig_11}. The distance between the secondary source and the error microphone was $0.2$ $\text{m}$, and the reference microphone array was placed $1.2$ $\text{m}$ from the error microphone. Acoustic paths were measured at several noise source locations\footnote[1]{Detailed path responses are available at \url{https://github.com/Wang-Boxiang/Spatial-Frequency-Generative-Fixed-Filter-Active-Noise-Control}}, while the positions of the reference microphones, secondary source, and error microphone remained fixed. Specifically, a \textit{Sennheiser AMBEO VR Mic} was used as the reference microphone array, which is consistent with the configuration employed in experiments with simulated RIRs. Consequently, the CRNN model trained on the simulated RIRs used in Section~\ref{Simulated Acoustic Paths} was directly applied to the real-world acoustic environment without additional retraining, while only the pre-trained control filter library was updated based on the measured acoustic paths.

Two experiments were conducted with different noise types and 3D locations. In the first experiment, the noise source was positioned at $(l = 0.47 \ \text{m},\phi  = 40^\circ, \theta  = 310^\circ )$ and emitted compressor noise. In the second experiment, the noise source was placed at $(l = 0.54 \ \text{m},\phi  = -20^\circ, \theta  = 250^\circ )$ and emitted hand drill noise. As shown in Table~\ref{table:CRNNoutputs}, the trained CRNN successfully classified the noise source into the nearest spatial classes in the pre-trained library. These results indicate that, when trained with sufficiently diverse data, the model exhibits a certain degree of robustness to real-world acoustic environments. 

\begin{figure}[!t]
\centering
\includegraphics[width=4in]{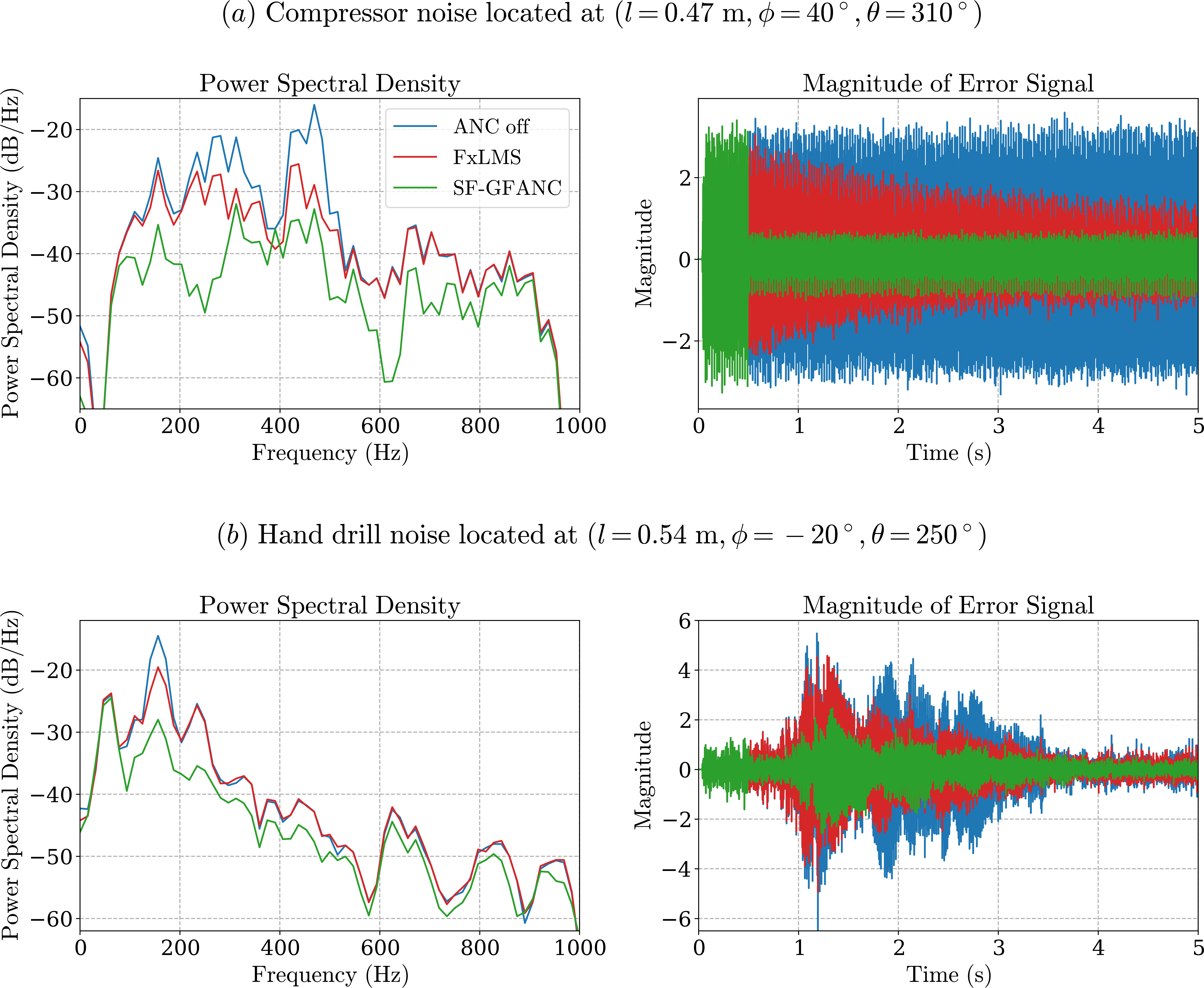}\vspace*{-0.4cm}
\caption{Noise reduction performance on real-world noises located at different 3D locations, evaluated with measured acoustic paths, using the FxLMS and the proposed SF-GFANC.}
\label{fig_12}
\vspace*{-0.3cm}
\end{figure}

Fig.~\ref{fig_12} presents the noise reduction performance of the SF-GFANC method in comparison with the FxLMS algorithm when attenuating compressor noise and hand drill noise. These results show that the SF-GFANC method achieves an NR of over $10$ dB for the main noise components. Furthermore, it demonstrates a significantly faster response compared to the FxLMS algorithm, which requires a longer convergence period and careful step size tuning to reach comparable performance, particularly under dynamic noise conditions.



\vspace*{-0.3cm}
\section{Conclusion}
\vspace*{-0.3cm}
\label{conclusion}
In this paper, we propose a novel SF-GFANC method that jointly exploits the noise source’s 3D spatial and frequency information to improve noise reduction performance in reverberant environments. Unlike the vanilla GFANC method, which only considers frequency characteristics, the proposed approach employs a multi-task CRNN to simultaneously estimate the source distance, elevation angle, azimuth angle, and the combination weights of sub control filters. By integrating both spatial and frequency cues, the system is able to accommodate noise sources with diverse 3D locations and frequency characteristics. In addition, the theoretical analysis formulates an explicit dependence of the optimal control filter on 3D spatial parameters under reverberant conditions.

Numerical simulations using both simulated and measured acoustic paths demonstrate that the proposed CRNN achieves accurate 3D localization and frequency estimation while generalizing well to unseen noise types and acoustic environments. Furthermore, comparative evaluations against representative ANC algorithms show that SF-GFANC consistently outperforms methods without 3D spatial information. Moreover, SF-GFANC provides a fast response, avoiding the slow convergence of adaptive ANC algorithms. Future work will focus on the real-time implementation and deployment of the SF-GFANC method.

\section*{Acknowledgment}
This work was supported by the Ministry of Education, Singapore, through Academic Research Fund Tier 2 under Grant MOET2EP50122-0018.


\bibliographystyle{elsarticle-num} 
\vspace*{-0.4cm}
\bibliography{refs}

\end{document}